\documentclass[10pt, conference]{IEEEtran}

\usepackage{subcaption} 
\usepackage{multirow}
\usepackage[utf8]{inputenc}

\usepackage{listings} 
\usepackage{xcolor}
\usepackage{amssymb}
\usepackage{wrapfig}

\usepackage{etoolbox}
\BeforeBeginEnvironment{appendices}{\clearpage}

\definecolor{codegreen}{rgb}{0,0.6,0}
\definecolor{codegray}{rgb}{0.5,0.5,0.5}
\definecolor{codepurple}{rgb}{0.58,0,0.82}
\definecolor{backcolour}{rgb}{0.95,0.95,0.92}

\lstdefinestyle{mystyle}{
    backgroundcolor=\color{backcolour},   
    commentstyle=\color{codegreen},
    keywordstyle=\color{magenta},
    numberstyle=\tiny\color{codegray},
    stringstyle=\color{codepurple},
    basicstyle=\ttfamily\footnotesize,
    breakatwhitespace=false,         
    breaklines=true,                 
    captionpos=b,                    
    keepspaces=true,                 
    numbers=left,                    
    numbersep=5pt,                  
    showspaces=false,                
    showstringspaces=false,
    showtabs=false,                  
    tabsize=2
}

\usepackage{booktabs} 
\usepackage{algorithm} 
\usepackage[noend]{algpseudocode} 
\usepackage{graphicx}
\algdef{SE}[DOWHILE]{Do}{doWhile}{\algorithmicdo}[1]{\algorithmicwhile\ #1}%
\algblockdefx{ParForAll}{EndParFor}[1]%
  {\textbf{for all }#1 \textbf{do in parallel}}%
  {\textbf{end for}}
\algtext*{EndParFor}

\algrenewcommand\textproc{}

\usepackage{graphicx}
\usepackage{color}
\usepackage{enumerate}
\usepackage{enumitem}
\usepackage{hyperref}
\usepackage{array}
\usepackage{listings}
\usepackage{verbatim}
\usepackage[frozencache=true,cachedir=.]{minted}
\usepackage{tcolorbox}
\usepackage{xcolor}
\usepackage[export]{adjustbox}

\usepackage{sidecap} 
\usepackage{amsmath} 
\usepackage{amsfonts}
\usepackage{cleveref}

\definecolor{darkblue}{rgb}{0,0,0}
\definecolor{darkred}{rgb}{0,0,0.0}

\providecommand{\linesref}[2]{\hyperref[#1]{Lines~\ref*{#1}--\ref*{#2}}}
\providecommand{\Linesref}[2]{\hyperref[#1]{Lines~\ref*{#1}} \hyperref[#2]{and~\ref*{#2}}}
\providecommand{\lineref}[1]{\hyperref[#1]{Line~\ref*{#1}}}
\providecommand{\Lineref}[1]{\hyperref[#1]{Line~\ref*{#1}}}

\usepackage[compact]{titlesec}
\newcommand{\tup}[1]{\left< #1 \right>}			

\newcommand{\func}[3]{#1 \colon #2 \rightarrow #3}	
\newcommand{\rem}[1]{}

\newcommand{\sub}{\subseteq}		

\renewcommand{\int}{\cap}
\newcommand{\define}{ \mathrel{\bf \colon \kern -2pt =} }	
\newcommand{\ints}{\mathbb{Z}}
\newcommand{\kskel}{k\mbox{-skel}}
\newcommand{\st}{ \mathrel{\colon} }	

\newcommand{\hg}{{\mathcal H}}

\newcommand{\adj}{\hbox{adj}}
\newcommand{\inc}{\hbox{inc}}

\newcommand{\bH}{{\bf H}}

\newcommand{\mypng}[3]{
	\begin{figure}[htbp]
	\begin{center}
	\includegraphics[scale=#1]{#2.png}
	\caption{\small #3}
	\label{#2}
	\end{center}
	\end{figure}
}

\usepackage[compact]{titlesec}
    \titlespacing{\section}{0pt}{1ex}{1ex}
    \titlespacing{\subsection}{0pt}{1ex}{0ex}
    \titlespacing{\subsubsection}{0pt}{0.5ex}{0ex}

\title{
High-order Line Graphs of Non-uniform Hypergraphs:
 Algorithms, Applications, and Experimental Analysis
}

\author{
  \IEEEauthorblockN{
  Xu T. Liu\IEEEauthorrefmark{1}\IEEEauthorrefmark{2},
  Jesun Firoz\IEEEauthorrefmark{3},
  Sinan Aksoy\IEEEauthorrefmark{3},
  Ilya Amburg\IEEEauthorrefmark{3},\\
  Andrew Lumsdaine\IEEEauthorrefmark{2}\IEEEauthorrefmark{3},
  Cliff Joslyn\IEEEauthorrefmark{3},
  Assefaw H. Gebremedhin\IEEEauthorrefmark{2},
  Brenda Praggastis\IEEEauthorrefmark{3}
  }
  \IEEEauthorblockA{
    \IEEEauthorrefmark{1}University of Washington,
    \IEEEauthorrefmark{2}Washington State University,
    \IEEEauthorrefmark{3}Pacific Northwest National Lab, USA
  }
  \IEEEauthorblockA{
    \IEEEauthorrefmark{1}\{x0, al75\}@uw.edu,
    \IEEEauthorrefmark{2}\{assefaw.gebremedhin\}@wsu.edu,
    \IEEEauthorrefmark{3}\{\{first name\}.\{last name\}\}@pnnl.gov
  }
}

\begin{document}

\maketitle

\begin{abstract}
Hypergraphs offer flexible and robust data representations for many applications, but methods that work directly on hypergraphs are not readily available and tend to be prohibitively expensive. Much of the current analysis of hypergraphs relies on first performing a graph expansion -- either based on the nodes (clique expansion), or on the edges (line graph) -- and then running standard graph analytics on the resulting representative graph. However, this approach suffers from massive space complexity and high computational cost with increasing hypergraph size. Here, we present efficient, parallel algorithms to accelerate and reduce the memory footprint of higher-order graph expansions of hypergraphs. Our results focus on the edge-based $s$-line graph expansion, but the methods we develop work for higher-order clique expansions as well.  To the best of our knowledge, ours is the first framework to enable hypergraph spectral analysis of a large dataset on a single shared-memory machine. Our methods enable the analysis of datasets from many domains that previous graph-expansion-based models are unable to provide. The proposed $s$-line graph computation algorithms are orders of magnitude faster than  state-of-the-art sparse general matrix-matrix multiplication methods, and obtain approximately $5-31{\times}$ speedup over a prior state-of-the-art heuristic-based algorithm for $s$-line graph computation.

\end{abstract}



\section{Introduction}
\label{sec:intro}

Hypergraph models are more natural representation than graphs for a broad range of systems---in biology, sociology, telecommunications, and physical infrastructures---involving {\em multi-way} relationships  \cite{Barabasi2016, berge1973graphs}, since graph models are limited to representing {\it pairwise} relationships.
Mathematically, a {\bf hypergraph} %
is a structure $\hg = \tup{V,E}$, with a set $V=\{v_j\}_{j=1}^n$  of vertices, and an indexable family $E = \{e_i\}_{i=1}^m$ of hyperedges $e_i \sub V$. Hyperedges have different sizes $|e_i|$, possibly ranging from the singleton $\{v\} \sub V$ (distinct from the element $v \in V$) to the vertex set $V$. 
A hyperedge $e=\{u,v\}$ with $|e|=2$ is the same as a graph edge. Indeed, all graphs $G = \tup{V,E}$  are hypergraphs: in particular, graphs are ``2-uniform'' hypergraphs, so that now $E \sub \binom{V}{2}$ and all $e \in E$ are unordered pairs with $|e|=2$. 
{\setlength{\abovedisplayskip}{3pt}
\setlength{\belowdisplayskip}{3pt}
An example hypergraph $\hg$ is shown %
in \Cref{fig:h_dual_2section} on vertices $V=\{a,b,\ldots,f\}$ and edges $ E = \{ 1:\{a,b,c\}, 2:\{b,c,d\},3:\{a,b,c,d,e\},4:\{e,f\} \}. $


A well-known method to study hypergraphs is to create a graph representation from the structure of the initial hypergraph using a graph expansion method such as the clique expansion~\cite{zien_multilevel_1999}.
The \textbf{clique expansion} replaces each hyperedge with a graph edge for each pair of vertices in the hyperedge. 
The information associated with hyperedges in the original hypergraph is lost in the new graph~\cite{KiS17}. Moreover, the size of the newly-constructed graph with these expansion methods increases exponentially~(\cite{jiang_hyperx_2019, heintz_mesh_2019}), which can significantly limit the scalability and applicability of these techniques. For example, there are approx. 10.3 billion edges in the clique-expansion graph of the Friendster dataset and 54.5 billion edges in that of Orkut~\cite{heintz_mesh_2019}. 
With billions of non-zero entries in the adjacency matrix of the clique-expansion graphs, processing these datasets is not possible 
on a single compute node. 

\begin{figure}[h]
\centering
\vspace{-1.6em}    
 {\includegraphics[scale=.4]{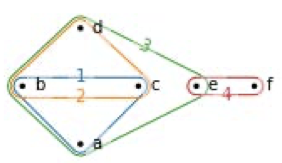}}
 {\includegraphics[scale=.4]{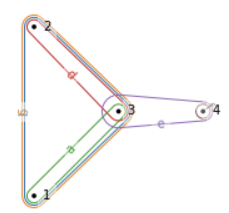}}
    \caption{\small \textit{(left) An example hypergraph $\hg$. (right) Dual $\hg^*$ of the example hypergraph $\hg$, defined later in \Cref{sec:background}.}}
\label{fig:h_dual_2section}
\vspace{-1.1em}
\end{figure}


In this work, we propose a scalable framework to study non-uniform hypergraphs with a lower-dimensional approximation of the original hypergraph called \textit{$s$-line graphs} of a hypergraph. Our multi-stage, versatile framework starts from the original hypergraph, and consists of multiple stages, including pre-processing, $s$-line graph construction, squeezing the $s$-line graph, and $s$-measure (defined later) computation. An $s$-line graph construction considers the number of common (overlapping) vertices, denoted by $s$, between each pair of hyperedges to capture the strength of connections among hyperedges. Such a model can represent, for example, the strength of the collaboration in a collaboration network. Specifically, we are interested in this work with only high-order $s$-line graphs, where $s\geq 2$. Compared with the clique-expansion graphs, the $s$-line graph of Friendster only has 53 edges and that of Orkut has 4,289 edges for $s=1024$. In an $s$-line graph, vertices (representing  hyperedges   of the  original  hypergraph) are connected when hyperedges intersect in at least $s$ hypergraph vertices in the original hypergraph.
\begin{wrapfigure}{l}{.2\textwidth}
 \vspace{-2ex}
 {\includegraphics[width=\linewidth]{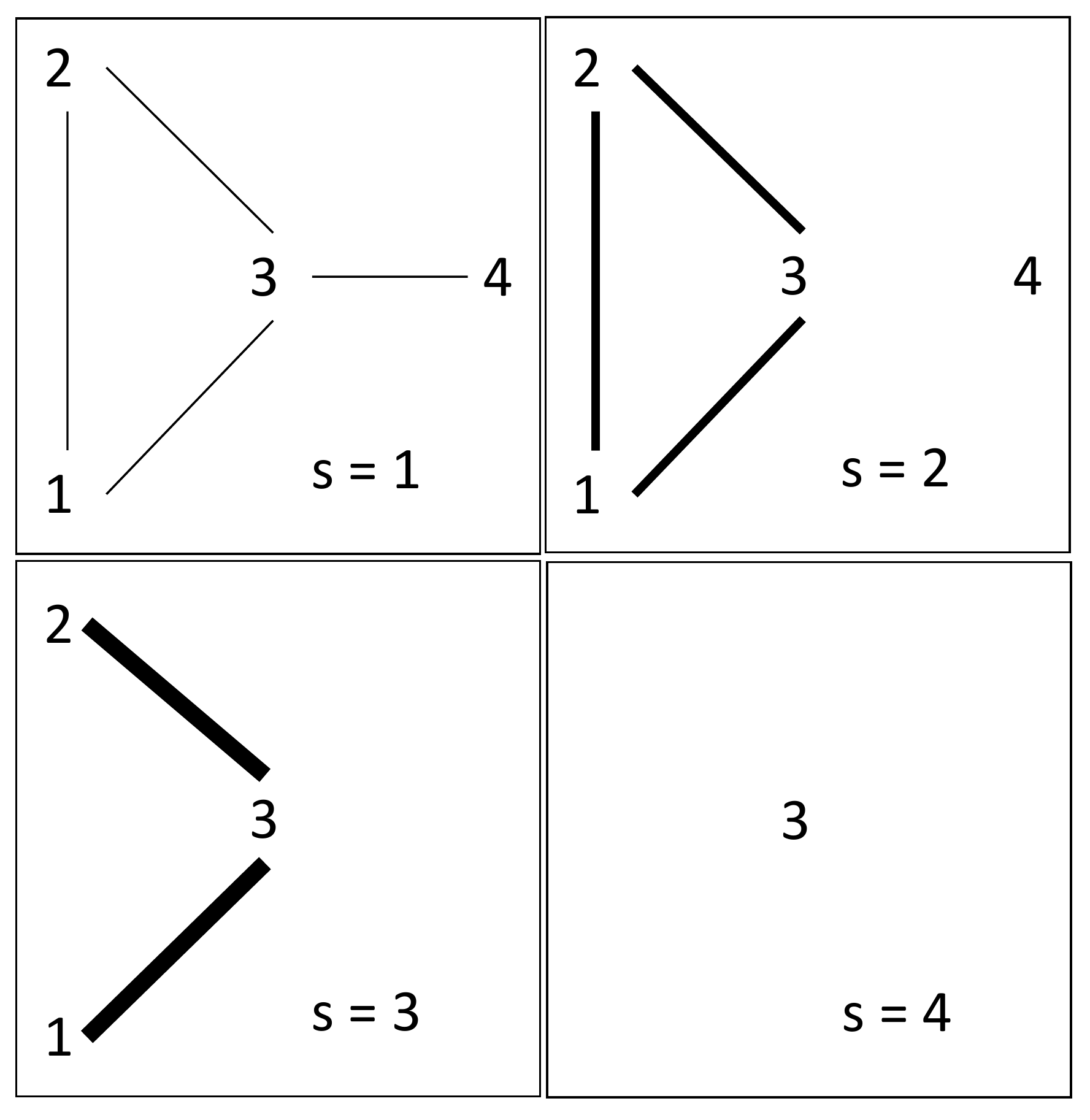}}
 \caption{\small \textit{Hyperedge $s$-line graphs $L_s(\hg) = \tup{E_s,F}$ for $s=1,2,3,4$ for the example in \Cref{fig:h_dual_2section}. The width of the graph edges represents the strength of the connection in the original hypergraph.}}
 \label{fig:slinegraphexample}
 \vspace{-2ex}
\end{wrapfigure} Dually, $s$-line graphs can also be constructed by considering the (hyper)vertices in the original hypergraph and their overlapping hyperedge sets. In this case, vertex $s$-line graph when $s=1$ is the clique-expansion graph of a hypergraph. \Cref{fig:slinegraphexample} shows the hyperedge $s$-line graphs $L_s(\hg)$ for our example for $s=1,2,3,4$. Note the changing vertex sets $E_s$ for each $s$ value, decreasing to $E_4{=}\{ 3 \}$ being the single hyperedge with $|e|{=}5 \ge 4$. Throughout this paper, we refer $s$-line graphs as hyperedge $s$-line graphs.

The drastic difference in size between the clique-expansion graphs and the s-line graphs has implications in the adjacency matrix representations of the graphs.
The size reduction entails drastic memory footprint reduction while computing a particular metric on the hypergraph (for example, when computing the Laplacian). Note that, in the $s$-line graph view of a hypergraph, as we vary the value of $s$, we can still retain the important connectivities in the original hypergraph. 

A naive approach for the $s$-line graph construction  is to find the intersection of the neighbor list of each pair of hyperedges in the original hypergraph. This is both compute- and memory-intensive. 
A recent parallel heuristic-based algorithm~\cite{firoz_2020_efficient} significantly improves the performance over the naive approach by avoiding redundant set intersections. However, the approach is based on heuristics and can only compute one $s$-line graph at a time with one $s$ value. \Cref{tab:comp_cost} compares 
the performance of the algorithm presented in \cite{firoz_2020_efficient} with the method proposed in this work in terms of runtimes on LiveJournal dataset. 
As observed from the table, the $s$-line computation stage is the most time-consuming step in the pipeline. Hence, we propose two new (exact) parallel algorithms for $s$-line graph construction to 
reduce the overall execution time and improve the efficiency of the process. We apply our framework to different datasets and real-world problems to gain insights into its performance and utility.

We identify three additional motivations for computing $s$-line graphs of a hypergraph. First, once computed, {\bf highly-tuned graph libraries} can be applied to the $s$-line graphs to measure different graph-theoretic metrics. The second motivation stems from {\bf applications}, where hypergraphs and $s$-line graphs enable 
new insights based on $s$-line graph metrics. 
Third, 
s-line graphs enable {\bf spectral graph analysis} of hypergraphs. To the best of our knowledge, there are no known method for directly computing the eigenvectors and eigenvalues of the rectangular incidence matrix of a hypergraph. The lack of a simple, eigenvalue-preserving algebraic relationship between the incidence matrix $H$ of a hypergraph, and the adjacency matrices of $s$-line graphs 
suggests the existence of a method for implicitly determining the s-line graph spectrum without forming the s-line graph itself is highly unlikely. Eigenvalues can provide insight into, for example, how well each of the connected components in an $s$-line graph remains connected and consequently provide insight about the original hypergraph connectivity. 

\begin{table}[t]
\small
\centering
\begin{tabular}{l|rr}
\hline
Stage                 & Algorithm in \cite{firoz_2020_efficient}       & our method             \\ \hline
preprocessing         & 0.122s          & 0.152s          \\
\textbf{s-overlap}     & \textbf{313.864s} & \textbf{12.085s} \\
squeeze               & 3.845s          & 2.656s          \\
s-connected components& 22ms   & 11ms  \\ \hline
total time            & 329.520s           & 28.216s           \\
\hline
speedup               & 1$\times$         & 26$\times$    \\
\hline
\#set intersections  & $8.66\times10^{9}$        & 0               \\ \hline
\end{tabular}
\vspace{0.5em}
\caption{\small \textit{Computational cost 
of each step of the high-order line graph framework 
 with the LiveJournal dataset~\cite{shun2020practical}. Clearly, $s$-overlap computation (in bold) is the dominant stage in the process. Note that our method does not perform any set intersection operation. }}\label{tab:comp_cost}
\vspace{0em}
\end{table}

{\bf Summary of contributions. } In this paper, we:
\vspace{-.2em}
\begin{itemize}[noitemsep,topsep=0pt]
    \item Propose two new hashmap-based $s$-line graph computation algorithms that completely avoid set intersection operations and prove to be significantly faster than the state-of-the-art efficient algorithm (\S\ref{sec:our_algos}).
    \item Propose a (C++ based) high performance, scalable framework for computing higher order line graph  of hypergraphs (\S\ref{sec:nwhy_framework}).
    \item Apply our framework on three real-world problems: uncovering collaborations in co-authorship networks and in
    co-staring networks, and identifying important genes in transcriptomics data. We demonstrate both higher efficiency and practical usability (\S\ref{sec:applications}).
    \item Empirically analyze scalability of our framework on a variety of real-world datasets and show superior performance over the algorithm proposed in~\cite{firoz_2020_efficient} (\S\ref{sec:expr}). We also compare our approach with a state-of-the-art sparse matrix-matrix multiplication (SpGEMM) library-based implementation (\S\ref{sec:spgemm-cmp}) and show superior performance.
\end{itemize}
\vspace{-.2em}

\section{Background}
\label{sec:background}

\subsection{Hypergraph Representations}

Hypergraphs may be represented in a number of equivalent forms. Given a hypergraph $\hg$, one can construct the {\bf bipartite graph} $B(\hg) = \tup{ V \sqcup E, E' }$ whose vertex set is the disjoint union of the hypergraph's vertices $V$ and hyperedges $E$, and whose edge set is the undirected graph edges $E' \sub {V \sqcup \binom{E}{2}}$, where $\{v,e\} \in E'$ iff $v \in e$. Further, one can  construct the Boolean {\bf incidence matrix} $\bH_{n \times m}$ where for $i \in [n], j \in [m]$, $b_{ij} = 1$ if $v_i \in e_j$, otherwise $b_{ij} = 0$. Note that $\bH$ is not square. 
These two representations are illustrated in Figure~\ref{fig:h_representation} for the example hypergraph introduced in \Cref{fig:h_dual_2section}. 

\begin{figure}[t]
\vspace{-0.5em}
    \begin{subfigure}[b]{0.49\linewidth}
    \centering
    \includegraphics[scale=.2]{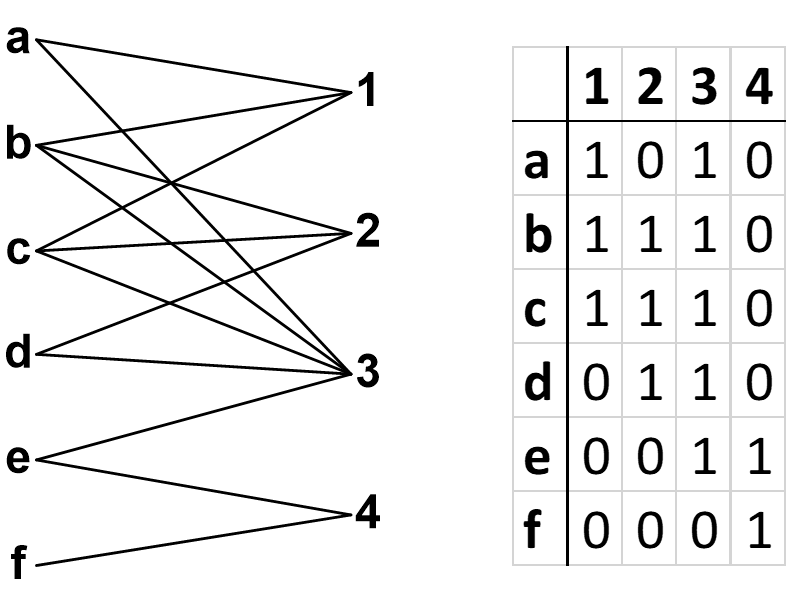}\label{figures/alternate}
    \end{subfigure}
    \begin{subfigure}[b]{0.49\linewidth}
    \includegraphics[scale=.5]{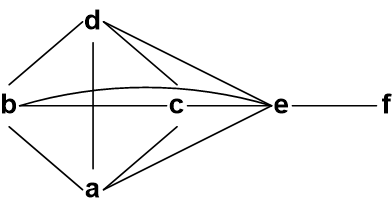}\label{figures/acda_2sec}
    \end{subfigure}
    \caption{\small \textit {(Left) Bipartite graph representation of $\hg$. (Middle) Incidence matrix $(\hg)$. (Right) 2-section $\hg_2$. \label{fig:h_representation}}}
\vspace{-1.1em}
\end{figure}

The {\bf dual hypergraph}  $\hg^*=\tup{E^*,V^*}$ of $\hg$ has vertex set $E^*=\{e_i^*\}_{i=1}^m$
and family of hyperedges $V^*=\{v_j^*\}_{j=1}^n$, where $v_j^* \define \{e_i^* : v_j \in e_i\}$. The dual $\hg^*$ for our example is shown in \Cref{fig:h_dual_2section}. 
$\hg^*$ is just the hypergraph with the transposed incidence matrix $H^T$, and $(\hg^*)^*=\hg$. 

In graphs, the structural relationship between two distinct vertices $u$ and $v$ can {\em only} be whether they are adjacent in a {\em single} edge ($\{u,v\} \in E$) or not ($\{u,v\} \not \in E$); and dually, that between two distinct edges $e$ and $f$ can {\em only} be whether they are incident at a {\em single} vertex ($e \cap f = \{ v \} \neq \emptyset$) or not ($e \cap f = \emptyset$). 
In hypergraphs, both of these concepts are applicable to {\em sets} of vertices and edges, and additionally become {\em quantitative}. Define $\func{\adj}{2^V}{\ints_{\ge 0}}$ and $\func{\inc}{2^E}{\ints_{\ge 0}}$, in both set notation and (polymorphically) pairwise:
{
\setlength{\abovedisplayskip}{0.2pt}
\setlength{\belowdisplayskip}{0.2pt}
   \[ \adj(U) = | \{ e \supseteq U \} |, \quad
		\adj(u,v) = | \{ e \supseteq \{ u,v \} \} |	\]
	\[ \inc( F ) = | \cap_{e \in F} e |,	\quad
		\inc(e,f)=| e \cap f |	\]
}
	for $U \subseteq V, u,v \in V, F \subseteq E, e,f \in E$.
	These concepts are dual, in that $\adj$ on vertices in $\hg$ maps to $\inc$ on edges in $\hg^*$, and {\it vice versa}. And for singletons, $\adj(\{v\}) = \deg(v) = |e \ni v|$ is the degree of the vertex $v$, while $\inc(\{e\}) = |e|$ is the size of the edge $e$. In our example, we have $\adj(b,c)=3$, while $\inc(\{1,2,3\})=2$. 
}

\subsection{Hypergraph Measures and \texorpdfstring{$s$}{}-Line Graphs}

Two edges $e,f \in E$ are
{\bf $s$-incident} if $\inc(e,f)=|e\int f| \ge s$ for $s \ge 1$.
\rem{An $s$-star is a set of edges $\mathcal{F} \subseteq \mathcal{E}$ sharing exactly a common
    intersection $f \subseteq V$, with $|f| \ge s$, so that $\forall e_i, e_j \in \mathcal{F}$ we have $e_i \cap e_j = f$.}
An {\bf $s$-walk} is a sequence of
edges $\tup{e_0,e_1,\ldots,e_n}$ such that each $e_{i-1},e_i$ are
$s$-incident for $1 \le i \le n$.
An $s$-path is an $s$-walk where no edges are repeated.

Aksoy {\it et al.}\ have developed  
various $s$-line graph metrics on the basis of $s$-walks~\cite{aksoy_hypernetwork_2020}. Here, we describe two of the metrics used in our paper. 
Let $E_s \define \{ e \in E \st |e| \ge s \}$.
The \textbf{$s$-betweenness centrality} of a hyperedge $e$ is $\sum_{f\neq g\in E_s} \frac{\sigma_{fg}^{s}(e)}{\sigma_{fg}^{s}} $, where $\sigma_{fg}^{s}(e)$ is the total number of shortest $s$-walks from hyperedge $f$ to $g$ and $\sigma_{fg}^{s}$ is the number of those shortest $s$-walks that contain hyperedge $e$.
A subset of hyperedges $F\subseteq E_s$ is an \textbf{$s$-connected component} if there is an $s$-walk between all edges $e, f \in F$, and $F$ is a maximal such subset.
These measures 
have important applications in hypernetwork science. 
For example, Feng {\it et al.}\ 
apply $s$-betweenness centrality to analyze biological datasets \cite{FeSHeE20}. 

Consider the {\bf $2$-section} $\hg_2 = \tup{V,F}$ of a hypergraph $\hg$ as a graph on the same vertex set $V$, but now with edges $F \sub \binom{V}{2}$ such that $\{u,v\} \in F$ iff there is some hyperedge $e \in E$ with $\{u,v\} \sub e$ (see Figure~\ref{fig:h_representation}). Thus $\hg_2$ can be thought of as a kind of ``underlying graph'' of a hypergraph $\hg$. 

Also of key interest is the 2-section of the dual hypergraph $\hg^*$, called the {\bf line graph} $L(\hg) = (\hg^*)_2$. Note that the vertices in $L(\hg)$ are the hyperedges in $E$, and two such (now) vertices $e,f \in E$ are connected with a line graph edge iff $\inc(e,f) > 0$. In general, for integer $s \ge 1$, define the {\bf $s$-line graph} of a hypergraph $\hg$ as a graph $L_s(\hg) = \tup{ E_s, F }$ where $F \sub \binom{E}{2}$ and $\{e,f\} \in F$ iff $e$ and $f$ are $s$-incident. It is known that in general, a hypergraph $\hg$ cannot always be reconstructed from  even {\em all} of the $s$-line graphs $L_s(\hg)$ {\em together with} the $s$-line graphs $L_s(\hg^*)$ of the dual \cite{KiS17}. Nonetheless, Aksoy {\it et al.}\ have demonstrated that all of the above measures {\em can} be calculated from the $s$-line graphs $L_s(\hg)$. 1-line graphs are also known as \textit{intersection graphs} or \textit{one-mode projections}.

\rem{there is substantial interest in understanding  exactly how much information about a hypergraph is available using only them. Line graphs are particularly widely used in studies when confronted with complex data naturally presenting as a hypergraph, despite the limitations.}

\newcommand{\bL}{{\bf L}}

$s$-line graphs can be naively calculated from the incidence matrix $\bH$, specifically, $\bL \define \bH^\top \bH$ is an $m \times m$ symmetric integer {\bf weighted adjacency matrix}, where each cell $\bL[i,j], i,j \in [m]$, records $\inc(e_i,e_j)$, and the diagonal entries $\bL[i,i]$ record edge size $\inc(\{e_i\}) = |e_i|$. For integer $s \ge 1$, define a Boolean filtration matrix $\bL_s$ where $\bL_s[i,j] = 1$ if $\bL[i,j] \ge s$, and 0 otherwise. Then $\bL_s - I$ is the adjacency matrix of $L_{s+1}$.

{
\setlength{\textfloatsep}{0pt}
\begin{algorithm}[t]
\caption{\small {Algorithm proposed in~\cite{firoz_2020_efficient} to compute 
the edge list of an $s$-line graph for a given $s$. \\
\textbf{Input:} Hypergraph $\mathcal{H} = (V, E)$, $s$ \\
\textbf{Output:} $s$-line graph edge list $L_s(\mathcal{H})$}}\label{algo:efficient_s_overlap_serial}
\begin{algorithmic}[1]
\State $L_s(\mathcal{H}) \gets \emptyset$
\State $L_{t}(\mathcal{H}) \gets \emptyset$, for each thread $t$
  \ParForAll{hyperedge $e_i \in E$}\label{efficient_serial:outterloop}
     \ForAll{vertex $v_k$ of $e_i$}\label{efficient_serial:neighofE}
        \ForAll{hyperedge $e_j$ of $v_k$ where ($i < j$)}\label{line:efficient_serial:neighofN}
            \State count $\gets  set\_intersection(neighbor\_list(e_i), \newline 
            neighbor\_list(e_j))$\label{line:eff_set_intersection}
            \If{count $\geq s $}
                \State $L_{t}(\mathcal{H}) \gets L_{t}(\mathcal{H}) \cup  \{e_i, e_j\}$
            \EndIf
        \EndFor
    \EndFor
  \EndParFor
  \State $L_s(\mathcal{H}) \gets L_s(\mathcal{H}) \cup $ every $L_{t}(\mathcal{H})$
\State \textbf{return} $L_s(\mathcal{H})$
\end{algorithmic}
\end{algorithm}

\section{Algorithms for Constructing \texorpdfstring{$s$}{}-line Graphs}
\label{sec:our_algos}
In this section, we start by briefly discussing a previous state-of-the-art algorithm for the $s$-line graph computation~\cite{firoz_2020_efficient} and derive the linear-algebraic equivalent formulation of the algorithm. We next transition to the linear algebraic formulation of our new algorithm and present our parallel $s$-line graph and ensemble $s$-line graph computation algorithms. Additionally, we discuss the design and implementation details of our parallel algorithms. We conclude the section with discussion about the distinctions between our algorithm and SpGEMM-based approach, the relationship of $s$-line graph with the weighted clique-expansion graph and the practicality of the s-line graph. Crucially, our methods also enable scalable analysis of higher-order clique expansions, but for the purpose of this work we mostly frame our language around, and present results for, $s$-line graph computations.   


\subsection{Previous Approaches}

Recently, Liu {\it et al.}\ proposed an algorithm~\cite{firoz_2020_efficient} (shown in ~\Cref{algo:efficient_s_overlap_serial}), where only the pair of hyperedges with at least one common neighbor is considered for the $s$-line graph computation. Additional heuristics have been applied to reduce the amount of redundant work. These heuristics include degree-based pruning, skipping already visited hyperedges, short-circuiting set intersection and considering either the upper or the lower triangular part of the adjacency matrix of the hyperedges. The proposed algorithm, in conjunction with these heuristics, achieves notable performance benefit over the naive approach.
While \Cref{algo:efficient_s_overlap_serial} 
improved the execution time of the $s$-line graph computation, performing explicit all-pairs set intersections despite incorporating different heuristics may still be computationally inefficient.

\subsection{Linear Algebraic Formulation of Our Algorithms}

Our approach exploits the linear algebraic relationships present in the adjacency matrix $\bL = \bH^\top \bH$. There are two basic variants to consider to construct $\bL$, which differ based on loop ordering.  In the first case, we consider the ``ijk'' loop ordering, where the inner loop is essentially a dot product between column $i$ and column $j$ of $\bH$, that is, an intersection between the non-zero locations of those two rows:  

\begin{algorithmic}[1]
\For {$i = 0, 1, \ldots $}
\For {$j = 0, 1, \ldots $}
\For {$k = 0, 1, \ldots$}
\State{$\displaystyle  \bL[i,j] \gets \bL[i,j] + \bH[k,i] \bH[k,j]$}
\EndFor
\EndFor
\State{$\displaystyle \bL_{s} \gets $ Boolean filtration on $\bL$ based on $s$}
\EndFor
\end{algorithmic}

An alternative ordering of the loops in matrix multiply interchanges the two inner loops.

\begin{algorithmic}[1]
\For {$i = 0, 1, \ldots $}
\For {$k = 0, 1, \ldots$}
\For {$j = 0, 1, \ldots $}
\State{$\displaystyle  \bL[i,j] \gets \bL[i,j] + \bH[k,i] \bH[k,j]$}
\EndFor
\EndFor
\State{$\displaystyle \bL_{s} \gets $ Boolean filtration on $\bL$ based on $s$}
\EndFor
\end{algorithmic}

In this case, the intersection is not so obvious.  The inner loop copies row $k$ of $\bH$, scaled by element $\bH[k,i]$, to row $i$ of $\bL$.  The ``intersection'' now is implicit in whether $\bH[k,i]$ is zero or non-zero. (In numerical linear algebra terminology, the inner loop is an ``axpy,'' or vector addition, operation.)

If we were to carry out this operation with actual matrices, the two forms would be computationally equivalent. However, we are carrying out this computation with graph structures, which are best represented as sparse matrices.  A computation using the graph structure, corresponding to the ``ijk'' ordering is given as
\begin{algorithmic}[1]
\For {$i = 0, 1, \ldots$}
\For {$j = 0, 1, \ldots$}
\State{$\bL[i,j] \gets \bL[i,j] + |  \bH.Adj[i] \cap 
\bH.Adj[j]|$}
\EndFor
\State{$\displaystyle \bL_{s} \gets $ Boolean filtration on $\bL$ based on $s$}
\EndFor
\end{algorithmic}
$\bH.Adj[i]$ indicates all vertices $k$ adjacent to vertex $i$ in $\bH$, so that $\adj(v_i,v_k)>0$.  Note that this form compares all pairs of vertices, which may be highly redundant if $\bH$ is sparse.

The alternative ``ikj'' formulation instead allows us to exploit the structure of the graph.
\begin{algorithmic}[1]
\For {$i = 0, 1, \ldots$}
    \For {$k \in \bH^{\top}.Adj[i]$} 
\For {$j \in \bH.Adj[k]$} 
\State{$\displaystyle  \bL[i,j] \gets \bL[i,j] + 1$}
\EndFor
\EndFor
\State{$\displaystyle \bL_{s} \gets $ Boolean filtration on $\bL$ based on $s$}
\EndFor
\end{algorithmic}
Here, rather than computing intersections between all pairs, we accumulate intersecting edges as we traverse the hypergraph.

{
\setlength{\textfloatsep}{0pt}
\begin{algorithm}[t]
\caption{\small {Our algorithm to compute 
the edge list of an $s$-line graph for a given $s$ using a hashmap data structure.\\
\textbf{Input:} Hypergraph $\mathcal{H} = (V, E)$, $s$ \\
\textbf{Output:} $s$-line graph edge list $L_s(\mathcal{H})$}}\label{algo:map_s_overlap_serial}
\begin{algorithmic}[1]
\State $L_s(\mathcal{H}) \gets \emptyset$
\State $L_{t}(\mathcal{H}) \gets \emptyset$, for each thread $t$
  \ParForAll{hyperedge $e_i \in E$}\label{line:map:outterloop}
  \If{$degree[e_i$] $< s$} \label{line:deg_prun_ei_start} \Comment{Degree-based pruning}
        \State \textbf{continue} 
  \EndIf \label{line:deg_prun_ei_end}
    \State overlap\_count $\gets []$ \label{line:map_decl}
     \ForAll{vertex $v_k$ of $e_i$}\label{line:map:neighofE}
        \ForAll{hyperedge $e_j$ of $v_k$ where ($i < j$)}\label{line:map:neighofN}
                \State overlap\_count[$e_j$]++ \label{line:map_overlap_count} \label{line:map:soverlap_store}
        \EndFor
    \EndFor
    \ForAll{$[e_j, n] \in $ overlap\_count}\label{map:edge_insert}
        \If{$n \geq s $}
            \State $L_{t}(\mathcal{H}) \gets L_{t}(\mathcal{H}) \cup  \{e_i, e_j\}$\label{line:map_insert}
        \EndIf
    \EndFor
  \EndParFor
  \State $L_s(\mathcal{H}) \gets L_s(\mathcal{H}) \cup $ every $L_{t}(\mathcal{H})$
\State \textbf{return} $L_s(\mathcal{H})$
\end{algorithmic}
\end{algorithm}
}

\subsection{Our Hashmap-based Algorithm to Compute a 
\texorpdfstring{$s$}{s}-line graph} 
\label{sec:map-based}

Based on the above observation, 
in contrast to performing an explicit set intersection between the full neighbor lists of both $e_i$ and $e_j$ (\Lineref{line:eff_set_intersection} in \Cref{algo:efficient_s_overlap_serial}), our new algorithm (\Cref{algo:map_s_overlap_serial}) only counts the common neighbor $v_k$ 
(\Lineref{line:map_overlap_count} in \Cref{algo:map_s_overlap_serial}). 
The new algorithm maintains a running count of the amount of overlaps between $e_i$ and $e_j$ observed so far. This is reminiscent of counting ``confirmed'' common members ($v_k$) between $e_i$ and $e_j$, instead of ``searching'' for common memberships between two neighbor lists of $e_i$ and $e_j$.  

To keep track of the running count, the algorithm allocates a hashmap data structure for each hyperedge $e_i$ (\Lineref{line:map_decl} in \Cref{algo:map_s_overlap_serial}) on the fly, with 2-hop neighbors $e_j$ as keys and the current overlap count of ($e_i, e_j$) as the values. The algorithm still considers only the set of edge pairs $(e_i, e_j)$ with at least one common neighbor ($v_k$) (\linesref{line:map:outterloop}{line:map:neighofN} in \Cref{algo:map_s_overlap_serial}) and these wedges are considered only from one direction $(i<j)$. 
We also apply degree-based pruning heuristic to filter out the set of hyperedges with degree $<s$ from the computation, as they are not members of $E_s$.


{
\begin{algorithm}[t]
\caption{\small {Our algorithm to compute 
the edge lists of an ensemble of $s$-line graphs using hashmap data structures.\\
\textbf{Input:} Hypergraph $\mathcal{H} = (V, E)$, $array\_s$ \\
\textbf{Output:} $s$-line graph edge lists $L_{s_i}(\mathcal{H}), \forall s_i\in array\_s $ 
}}\label{algo:map_s_overlap_ensemble_serial}
\begin{algorithmic}[1]
\State overlap\_count $\gets \{\}$
\State $s \gets$ smallest $s \in array\_s$
\ParForAll{hyperedge $e_i \in E$}\label{ensemble:outterloop}
    \If{$degree[e_i$] $< s$}
        \State \textbf{continue} 
    \EndIf
    \State overlap\_count[$e_i$] $\gets []$
    \ForAll{vertex $v_k$ of $e_i$}\label{ensemble:neighofE}
        \ForAll{hyperedge $e_j$ of $v_k$ where ($i < j$)}\label{line:ensemble:neighofN}
                \State overlap\_count[$e_i$][$e_j$]++ \label{line:ensemble_map_count}
        \EndFor
    \EndFor
\EndParFor
\ParForAll{$s_i \in array\_s$}\label{line:ensemble:edge_insert}
    \State $L_{s_i}(\mathcal{H}) \gets \emptyset$
    \ForAll{hyperedge $e_i \in E$}
        \ForAll{ $[e_j, n]  \in $overlap\_count[$e_i$]}
            \If{$n \geq s_i $}
                \State $L_{s_i}(\mathcal{H}) \gets L_{s_i}(\mathcal{H}) \cup  \{e_i, e_j\}$ \label{line:ensemble_create_edgelist}
            \EndIf
        \EndFor
    \EndFor
\EndParFor
\State \textbf{return} $L_{s_i}(\mathcal{H}), \forall  s_i \in array\_s$
\end{algorithmic}
\end{algorithm}
}

\vspace{-0.3em}
\subsection{Computing Ensemble of \texorpdfstring{$s$}{s}-line Graphs}
Occasionally, it may be required to compute an ensemble of $s$-line graphs, instead of a single one, for different values of $s$. In this scenario, running \cref{algo:map_s_overlap_serial} multiple times to generate $s$-line graphs separately may be inefficient. Hence, to compute an ensemble of $s$-line graphs, we modify \cref{algo:map_s_overlap_serial} to first accumulate and store the overlap counts, and then filter out edge-pairs based on a particular $s$ value. The modified algorithm is shown in \Cref{algo:map_s_overlap_ensemble_serial}. Since multiple $s$-line graph will be constructed, instead of the in-place insertion of edges ($e_i, e_j$) with $s$ overlapping neighbors in the $s$-line graph's edge list (\Lineref{line:map_insert} in \cref{algo:map_s_overlap_serial}), we decouple this insertion step from the counting step. The algorithm maintains a running count of overlaps for each pair of hyperedges ($e_i, e_j$)
(\Lineref{line:ensemble_map_count} in \cref{algo:map_s_overlap_ensemble_serial}). Once the counting step is completed, for each value of $s$, the algorithm loops through the hashmap containing all ($e_j,counts$) pairs for each $e_i$ and construct the edge list of the $s$-line graph (\linesref{line:ensemble:edge_insert}{line:ensemble_create_edgelist} in \Cref{algo:map_s_overlap_ensemble_serial}). Degree-based pruning can be applied 
to filter out the hyperedges with degree smaller than the smallest $s$ in $array\_s$. To avoid duplicate counting for a pair of edges $(e_i, e_j)$, we prune redundant computation related to edge $(e_j, e_i)$. 
%
%
%

\subsection{Parallel Time Complexity Analysis}
We analyze the complexity of \Cref{algo:map_s_overlap_serial} and \Cref{algo:map_s_overlap_ensemble_serial} in the work-depth model~\cite{jaja_intro_1992}. 
The work $W$ is equal to the total number of independent computations. The depth $D$ is equal to the time required for the critical path computation (in the computation DAG, the longest chain of dependency). If $P$ processors are available, with a randomized work-stealing scheduler, Brent's scheduling principle dictates that the running time is $O(W/P + D)$. Each hyperedge is visited once on the outermost loop ($|E|$). Without considering any heuristics, the second inner loop visits $\overline{d}_v$ number of incident hypernodes on average. The innermost loop visits $\overline{d}_e$ incident hyperedges on average. Because lookup and insertion of elements in a hashmap is constant on average, therefore, \Cref{algo:map_s_overlap_serial} takes $O(|E|\overline{d}_v\overline{d}_e)$ on average, and $O(|V||E|^2)$ time in the worst case. 
The overall work is $O(|V||E|^2)$, and overall depth is $O(log|H|)$. Here $|H|$ denotes the number of non-zero entries in the hypergraph incidence matrix.
\Cref{algo:map_s_overlap_ensemble_serial} has the same time complexity as \Cref{algo:map_s_overlap_serial}.
Next we consider degree-based pruning 
and considering only the upper triangular part of the adjacency matrix $\bL_s (e_i, e_j$ pairs with  $i<j)$. The degree-based pruning trims the work in outermost loop to $E_s$. 
Considering only the upper triangle of the adjacency matrix $\bL_s$ essentially cuts the overall work by half. 

\subsection{Parallel Implementation Design Considerations} 
We implement our framework in \texttt{C++17}. Since $s$-line graph computation is the most compute-intensive stage in the pipeline, we parallelize our algorithms to compute the $s$-line graphs in Stage 3. For this purpose, we leverage the parallel constructs available in Intel oneAPI Threading Building Blocks (oneTBB)~\cite{tbbrepo}. In particular, the outermost for loops iterating over the hyperedges in \Cref{algo:map_s_overlap_serial} and \Cref{algo:map_s_overlap_ensemble_serial} are parallelized with the  \texttt{parallel\_for} construct in oneTBB. \texttt{parallel\_for}, in the form of \texttt{(range, body, partitioner)}, allows different ranges to be passed in to enable partitioning the range (hyperedges) in different ways so that  different workload distribution strategies among the threads can be tested, as long as the provided range meets the \texttt{C++} range requirements. 


\textbf{Ranges and Partitioning strategies.} oneTBB provides a built-in range, namely \emph{blocked range}, where the hyperedges (IDs) can be divided into blocks (chunks) and each chunk of contiguous hyperedges (IDs) can be assigned to one thread. Additionally, we adopt an alternative, customized range, namely \emph{cyclic range}. 
Here, given the stride size equal to the number of total threads $nt$, thread 0 processes hyperedges $e_0, e_{0+nt}, e_{0+2*nt}, e_{0+3*nt}$ and so on, thread 1 processes hyperedges $e_1, e_{1+nt}, e_{1+2*nt}, e_{1+3*nt}$ and so on. Here $e_i$ denotes a hyperedge ID. 
oneTBB is based on work-stealing runtime scheduler. Work stealing scheduler is particularly beneficial in our context, since this enables idle threads to steal work from other straggler threads, which are currently processing, for example, high-degree hyperedges.

\textbf{Granularity Control.} To accommodate flexibility for load balancing, oneTBB also provides provision for specifying the \textit{granularity} of work done by each thread, while reducing the overheads of work stealing and task scheduling. We leverage this fine-grained control to specify the block size of the chunk of work (i.e. the number of hyperedges assigned to each thread). We notice that chunk size up to 256 achieves similar performance. With larger chunk sizes, the scheduling overhead noticeably impacts algorithm performance.

\textbf{Data Structures for the Main Performance Criterion (Overlap Count).} The hashmap data structures for maintaining the overlap\_counts in our algorithms are thread-local data structures, implemented with the C++ \texttt{ std::unordered\_map}. In \Cref{algo:map_s_overlap_ensemble_serial}, for example, each hyperedge is associated with a hashmap that maintains a list of neighbors with at least one overlapping vertex. Before applying filtering (s), the size of each of these individual hashmap is equal to the degree of each hyperedge. With hypergraphs with skewed-degree distribution, s-line computation may have hashmaps for which the sizes vary significantly.

\textbf{Consideration of dynamic vs pre-allocated thread-local storage:} We have observed that pre-allocated thread-local storage (TLS) (i.e. per-thread hashmap allocated outside of the outermost for loop and resetting it after each iteration) may be beneficial for computing $s$-line graphs with hypergraphs with denser overlapping neighbor sets for each pair of hyperedges. Web dataset, discussed in \Cref{sec:expr}, is one such example. For a particular $s$ value, Web generates denser $s$-line graph. Dynamically allocating and deallocating a hashmap in each iteration on-the-fly inside the outermost for loop is costlier in this case. All other datasets, however, prefer dynamically-allocated hashmap for each thread in each iteration.




\subsection{Relationship among 
Our Hashmap-based \texorpdfstring{$s$}{s}-line Graph Algorithm, \Cref{algo:efficient_s_overlap_serial} 
and Sparse Matrix-Matrix Multiplication (SpGEMM).}

When constructing a single $s$-line graph for a particular $s$ value, considering the pairs of hyperedges sharing at least one common node is equivalent to computing the sparse general matrix-matrix multiplications (SpGEMM)~\cite{gustavson_1978_two} followed by a filter operation to find the edgelist of an $s$-line graph. However, the SpGEMM-based approach is both time-consuming and memory-intensive. There are three reasons why it is not efficient for computing $s$-line graphs. First, it considers both the upper triangular and the lower triangular of hyperedge adjacency matrix $\bL_s$ 
even though the matrix is symmetric. In contrast, our algorithm can exploit this symmetry to consider either the upper or the lower triangular part of the matrix. Second, since SpGEMM is more general, it has to compute and store the product matrix before applying filtration upon the matrix. This requires extra space to store the intermediate results (i.e., the product matrix). Our algorithm, on the other hand, can apply the filtration operation on-the-fly and does not require to materialize the product matrix due to the known $s$ value.  Third, the SpGEMM-based approach cannot apply other heuristics to speedup the computation, such as degree-based pruning (prune all the hyperedges with degree $< s$) or short circuit the set intersection as applied in~\Cref{algo:efficient_s_overlap_serial}. We report the performance comparison 
of our algorithms with a state-of-the-art parallel SpGEMM library 
in \Cref{sec:spgemm-cmp}. 

\subsection{Relation to the (Weighted) Clique-expansion Graph}\label{subsec:clique_expansion_rel}
Given a hypergraph $\hg$ with incidence matrix $\bH$, we can compute the weighted clique-expansion adjacency matrix as ${\bf W}=\bH\bH^T-{\bf D}_V$ where ${\bf D}_V$ is a diagonal matrix with node degrees as its diagonal entries.  It is easy to see that ${\bf W}[i,j]$ is the number of hyperedges nodes $i$ and $j$ appear together in. Note that we can use ${\bf W}$ to obtain $L_s(H^{*})$ for every integer $s\geq 1$ through its adjacency matrix $\bL^{*}_s.$ We set $\bL^*_s[i,j]=1$ if ${\bf W}[i,j]\geq s$ and 0 otherwise. However, the above procedure would be very memory-intensive as ${\bf W}$ can be very dense. 
 
This observation implies that we could use our approach to efficiently compute $s$-sections, or ``$s$-clique'' graphs, where a graph edge connects two nodes if the nodes appear together in a hyperedge at least $s$ times, bypassing memory limitation issues by not having to explicitly compute ${\bf W}$. In particular, this could be accomplished by running our algorithm to directly compute $L_s(\bH^*)$ for a given $s$. So in other words, the $s$-line graph problem is dual to the $s$-clique problem. Although we frame our paper through the $s$-line graph perspective, it is crucial to note that the tools we develop apply equally well to the $s$-clique graph problem. The choice of which perspective to take depends on whether one wants to investigate edge- ($s$-line graph) or node- ($s$-clique graph) centric properties, and on the particular application. 

\rem{
 using linear algebraic approaches are sound to calculate

We can derive formulae for computing an s-line graph  
$G = \langle U, F\rangle = \langle E, F\rangle$
from a hypergraph ${\cal H} = \langle V, E \rangle$
with the following observation.  Given the set of
hyperedges 
\(
e_i \in E, i = 0, 1, \ldots
\)
we define a set of vertices 
\(
u_i \in U, i = 0, 1, \ldots
\)
where each $u_i$ corresponds to $e_i$.  We have an edge between
two vertices $u_i$ and $u_j$ if there is a non-empty intersection between the
sets $e_i$ and $e_j$, i.e., 
$\{u_i, u_j\} \in F \leftrightarrow 
e_i \cap e_j \neq \varnothing $

We can consider algorithms for computing $L_s$ based on the structure of matrix-matrix product according to 
\[
L = H H^\top - 2 I .
\]

}

\subsection{Motivation for using higher-order graph expansions}
A widespread approach to hypergraph analysis is to focus instead on associated graph projections, such as the clique expansion. As discussed in \Cref{subsec:clique_expansion_rel}, our framework actually includes the clique expansion as a special case: the $s$-line graph of the dual hypergraph (i.e. $s$-clique graph) is the graph obtained by linking vertices in the hypergraph whenever they belong to $s$ or more shared hyperedges. In this way, the $1$-line graph of the dual hypergraph is the clique expansion. Compared to the clique expansion approach, there are significant, practical benefits afforded by the $s$-clique approach, for $s>1$.

In particular, $s$-clique graphs can reduce the density of graph projections while preserving -- or even amplifying -- essential features of the network. Line graphs (or clique expansions) of hypergraph-structured data tend to be prohibitively dense because a single high degree vertex (resp., large hyperedge) yields quadratically many edges. For instance, in an author-paper hypergraph, a single paper with many authors (i.e. large hyperedge) links all pairs of those authors, whereas for $s>1$, the $s$-clique graph approach requires {\it more than one} joint paper to link those authors in the collaboration graph. 


\begin{figure}
    \centering
    \includegraphics[width=0.8\linewidth]{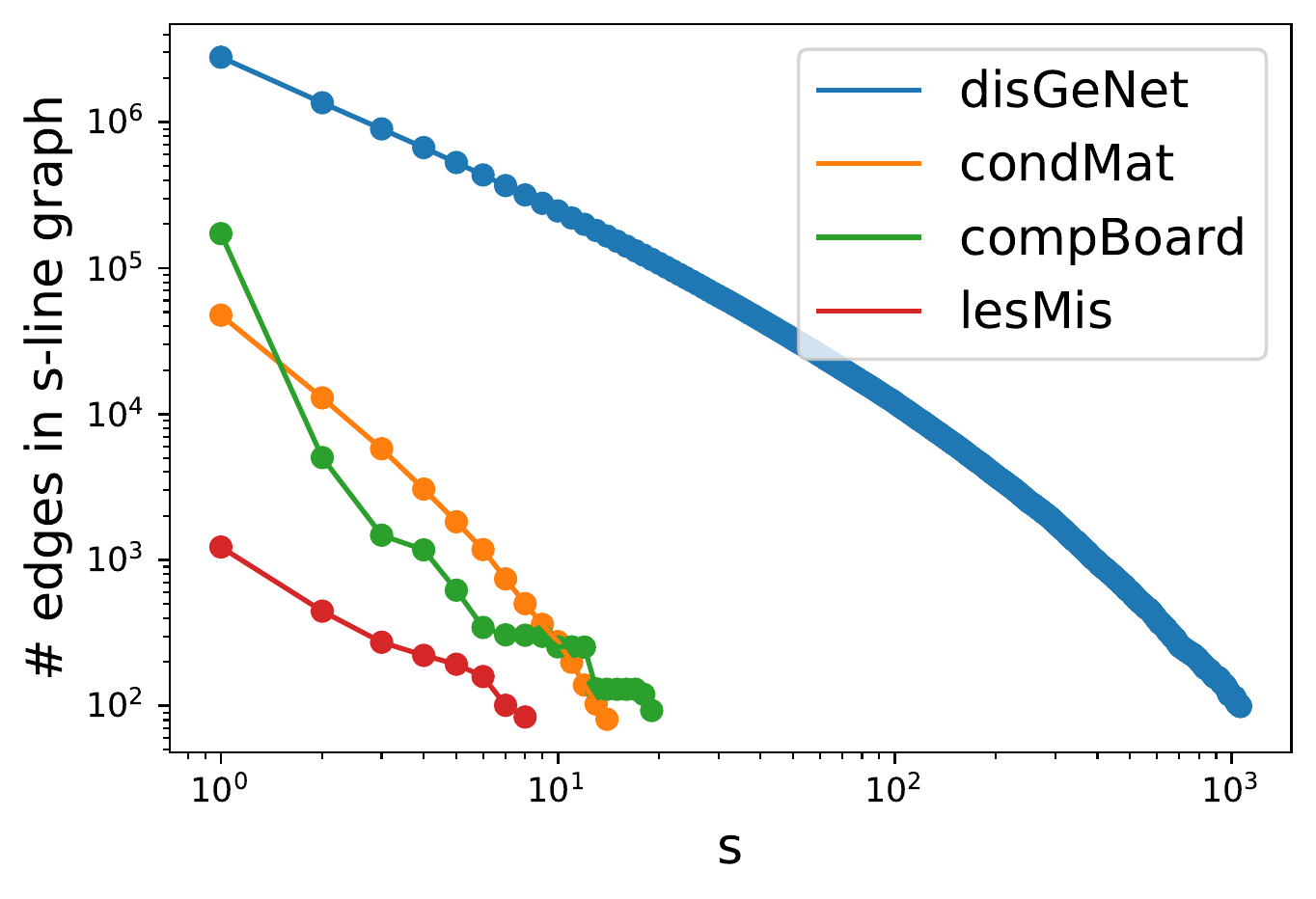}
    \caption{\small \textit{The number of edges in the $s$-clique graph of four datasets}}
    \label{fig:density}
\end{figure}

In practice, we find the density of $s$-clique graphs drops off exponentially in $s$ in data sets from far-ranging domains. 
In log-log scale, Figure \ref{fig:density} plots the number of edges in $s$-clique graphs against $s$ for disGeNet (a disease-gene dataset
~\cite{pinero2020disgenet}), condMat (an author-paper network from the condensed matter section of the arXiv \cite{structure_newman_2001}), compBoard (a board member-company network from \cite{aksoy_hypernetwork_2020}), and lesMis (a character-scene network derived in \cite{knuth1993stanford} from Victor Hugo's Les Miserables). While the rates of decrease differ across datasets, $s$-clique graphs rapidly sparsify as $s$ increases. For larger datasets, the formation of the clique expansion is intractable; $s$-clique graphs provide an alternative in these cases. 

\begin{table}
\begin{center}
{\begin{tabular}{l|lll} 
 \hline
 \multirow{2}{*}{Disease} & \multicolumn{3}{c}{Rank \& Score Percentile} \\
 & $s=1$ & $s=10$ & $s=100$ \\ 
 \hline
 Malignant neoplasm of breast & 1 (100\%)& 1 (100\%) & 1 (100\%) \\ 
 Breast carcinoma & 2 (99.99\%) & 2 (99.99\%) & 2 (99.99) \\
 Malignant neoplasm of prostate & 3 (99.97\%) & 4 (99.96\%) & 4 (99.96\%) \\
 Liver carcinoma & 4 (99.96\%) & 3 (99.97\%) & 3 (99.98\%) \\
 Colorectal cancer & 5 (99.95\%) & 5 (99.95\%) & 6 (99.94\%) \\
 \hline
\end{tabular}}
\end{center}
\caption{\small \textit{Ordinal rank and score percentile of the top 5 diseases by PageRank score in the clique expansion (i.e. $s=1$), as well as the $s$-line graphs of the dual hypergraph ($s$-clique expansion), for $s=10,100$.}}\label{tab:ranks}

\end{table}

Even when $s$-clique graph formation is feasible for $s=1$, focusing on $s>1$ may be sufficient or preferable for a number of basic analytic tasks. 
While this of course is data and question dependent, we illustrate the potential effectiveness of this approach for one common analytical task: centrality and ranking. In biology, hypergraphs have been utilized to identify structurally critical genes and diseases in interactome networks \cite{goh2007human}.
Returning to the disease-gene network, we construct the clique expansion (linking diseases associated with common genes), compute the PageRank score of the diseases, and compare this to the PageRank rankings of diseases in the $s$-clique graphs, for $s=10$ and $s=100$. Table \ref{tab:ranks} presents how the top 5 ranked diseases in the clique expansion ($s=1$) are ranked in the $s=10$ and $s=100$ higher-order clique expansions. These three graphs are of vastly different densities, having 2.7M, 246K, 12K edges, respectively. Nonetheless, the ordinal rankings and score percentiles for the top 5 rated diseases are nearly identical across all three graphs. Extending to the top 400 diseases -- which constitute those above 95\% percentile of scores -- shows that $92\%$ and $88\%$ of these diseases remain in the top 400 for $s=10$ and $s=100$, respectively. In this case, the higher-order $s$-clique graph approach identifies essentially the same critical diseases according to their PageRank using a network with 231 times fewer edges than the clique expansion.






\section{Our \texorpdfstring{$s$}{s}-line Graph Computation Framework }
\label{sec:nwhy_framework}

We now discuss our $s$-line graph framework for non-uniform hypergraphs in detail.
The framework has five major stages, two of which are at least partially optional, depending on the needs of a particular data set and  problem.




\textbf{Stage-1\quad Pre-processing.} Pre-processing hypergraph includes removing isolated vertices, empty edges, and relabeling.

    
\textbf{Relabeling.} Large hypergraphs with highly-skewed, non-uniform degree distributions generally benefit from relabeling the hyperedge IDs according to their degrees (henceforth referred to as \emph{relabel-by-degree}). 
    Let's consider a ``wedge'' motif $(e_i, v_k, e_j)$ in the bipartite graph hypergraph form $B(\hg)$. When counting the common neighbor $v_k$, to avoid considering $v_k$ twice: once in view of ($e_i, v_k, e_j$) and another as ($e_j, v_k, e_i$), all $s$-line computation algorithms include a comparison ($i<j$), so that the ``wedge'' is traversed only once (\Lineref{line:efficient_serial:neighofN} in \Cref{algo:efficient_s_overlap_serial}, \Lineref{line:map:neighofN} in \Cref{algo:map_s_overlap_serial} and \Lineref{line:ensemble:neighofN} in \Cref{algo:map_s_overlap_ensemble_serial}). This is equivalent to considering only the upper triangle of the adjacency matrix $\bL_s$. 
    
    Relabel-by-degree in ascending order, in conjunction with considering the upper triangle of $\bL_s$, may improve the performance of the algorithm. Additionally, this helps achieve better load balancing among threads while executing a parallel $s$-line graph computation algorithm in the later stage.  Equivalently, relabel-by-degree in {\em descending} order, in conjunction with considering the {\em lower triangle} of $\bL_s$, may provide similar performance improvement.


\textbf{Stage-2 (optional) 
Computing toplexes.} We calculate the toplexes  $\check{E}$, and thereby the simplified hypergraph $\check{\hg}$. A {\bf toplex} is a maximal edge $e$ such that there exits no edge $f$ where $\not\! \exists f \supseteq e$. Let $\check{E} \sub E$ be the set of all toplexes. 
For a hypergraph $\hg$,  $\check{\hg} = \tup{V,\check{E}}$ is the {\bf simplification} of $\hg$, and  $\hg$ is {\bf simple} when $\hg = \check{\hg}$, 
so that all edges are toplexes. A simplification may result in significantly smaller $\check{\hg}$, which may, in turn, reduce the memory footprint of subsequent stages. 
Efficient  algorithms for computing toplexes  \cite{marinov2016practical} are available. 


\vspace{-0.1em}
\textbf{Stage-3\quad Computation of the edgelist of the $s$-line graph of a given hypergraph.} The most important and compute-intensive stage of the $s$-line graph framework involves construction of the $s$-line graph itself. Depending on the requirement, the objective of this stage can be two-fold: the computation of \emph{only one} $s$-line graph for a particular $s$ value or an \emph{ensemble} of $s$-line graphs for different values of $s$. Computation of an ensemble of $s$-line graphs is more memory-intensive in comparison to just computing a single $s$-line graph. We discuss in detail two algorithms for computing individual and ensemble of line graphs in the next section. 

\vspace{-0.1em}
\textbf{Stage-4\quad ID squeezing (optional) and $s$-line graph construction.} After we finish computing the edgelist of the $s$-line graphs, many hyperedge pairs may not be included in the newly-constructed $s$-line graph due to insufficient overlap between their vertex sets. Hence, the adjacency matrix of the $s$-line graph may be \textit{hypersparse} (many rows will be empty when considering $s$-overlap). Retaining the original IDs of the edges to construct the new $s$-line graph will thus be wasteful in terms of memory. Hence, optionally, we may remap the IDs to a contiguous space to eliminate the ``holes'' in the ID space of the $s$-line graph. This stage is called \emph{ID squeezing}. 
The $s$-line graph is constructed based on the generated edgelist.

\vspace{-0.1em}
\textbf{Stage-5\quad $s$-metric computation.} Once the $s$-line graph is constructed, different $s$-line graph metrics are computed, including $s$-connected components, $s$-centrality, $s$-distance, etc. When computing these metrics, any standard, relevant graph algorithm can be applied to compute such metrics. 

\begin{figure}[ht]
\vspace{-0.5em}
     \centering
\begin{adjustbox}{minipage=\columnwidth,scale=0.7}
    \begin{subfigure}[b]{0.3\linewidth}
       \centering
       \includegraphics[width=\linewidth]{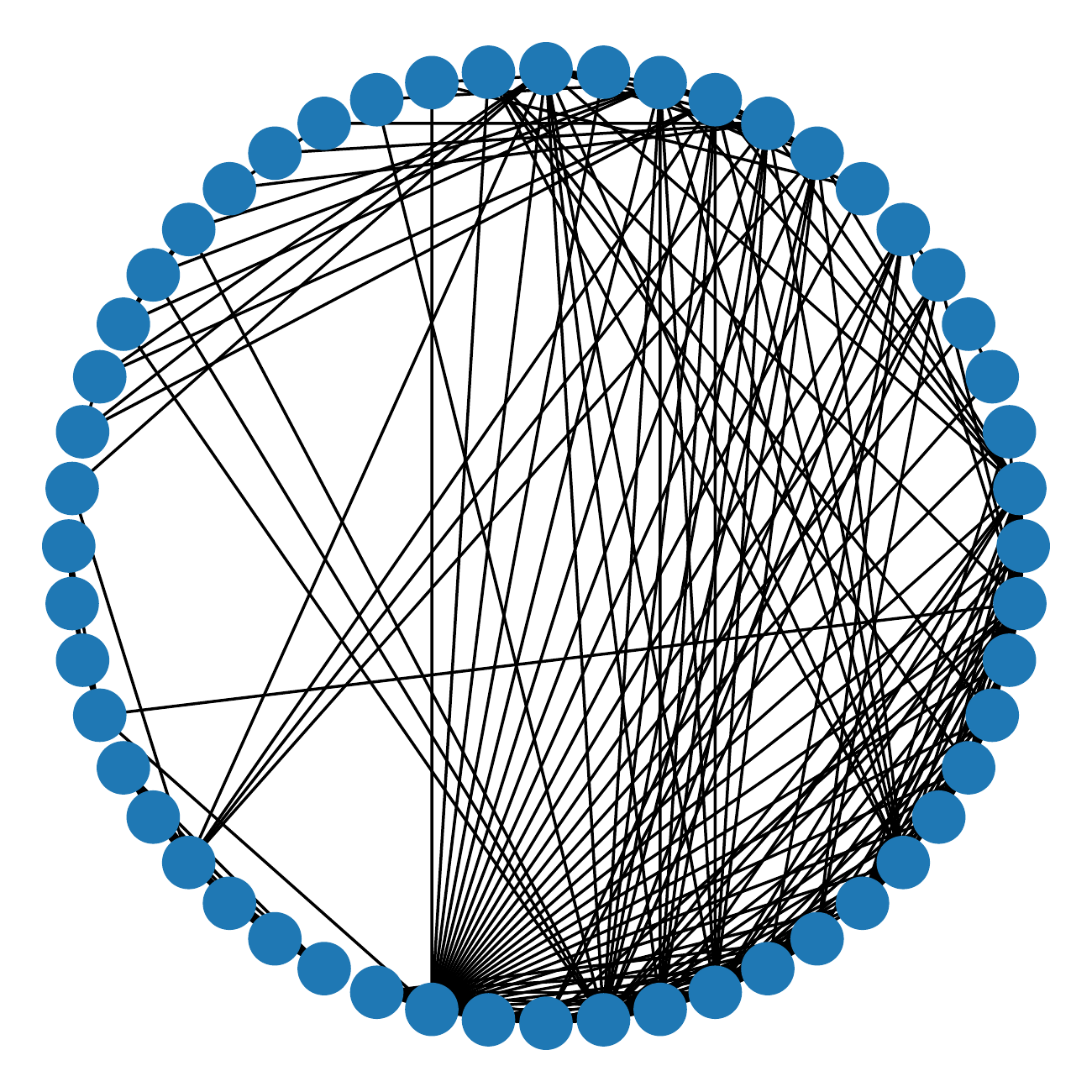}
        \caption{$s=1$}
        \label{fig:1lineggraph}
    \end{subfigure}
~
    \begin{subfigure}[b]{0.3\linewidth}
       \centering
       \includegraphics[width=\linewidth]{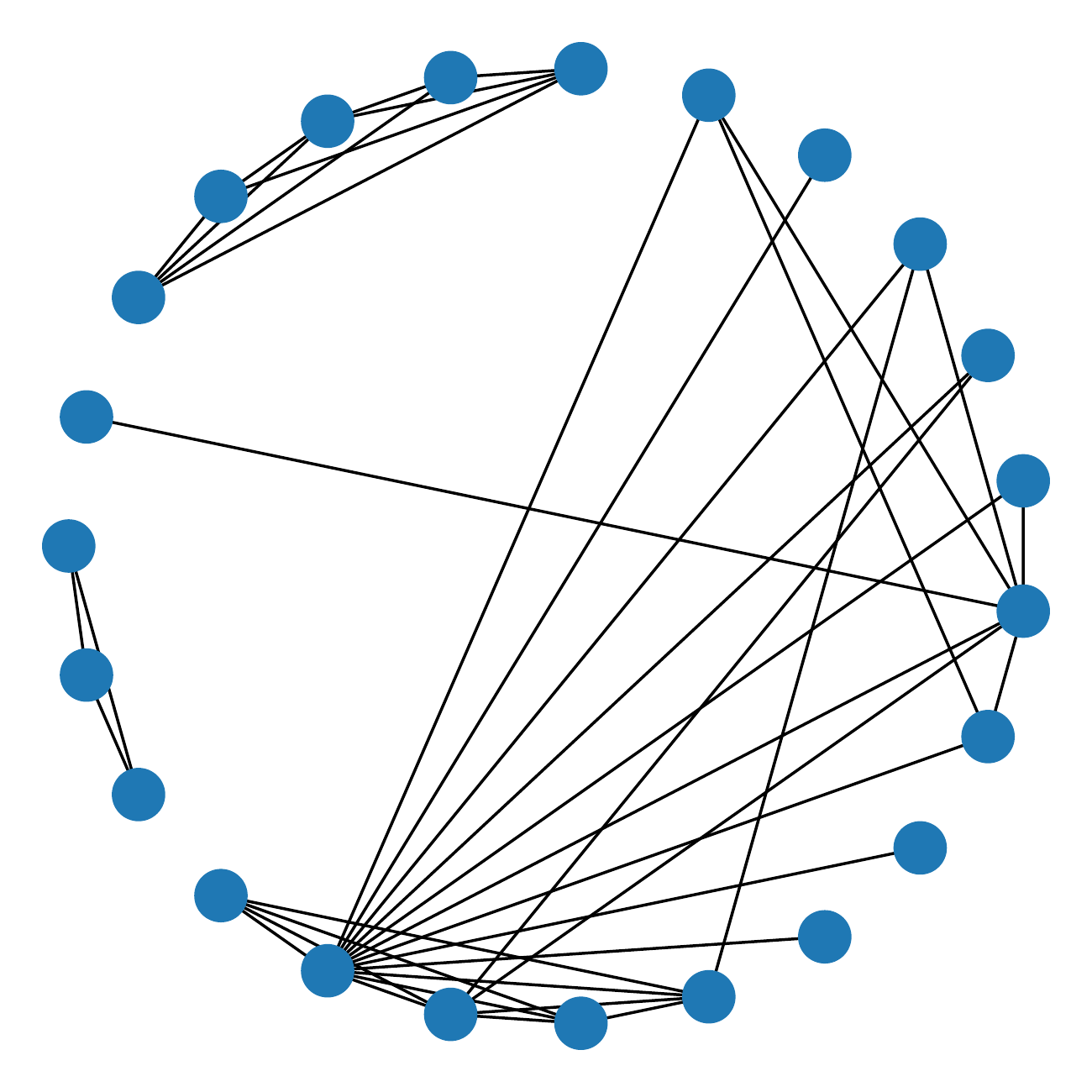}
        \caption{$s=3$}
        \label{fig:3linegraph}
    \end{subfigure}
    ~
    \begin{subfigure}[b]{0.3\linewidth}
       \centering
       \includegraphics[width=\linewidth]{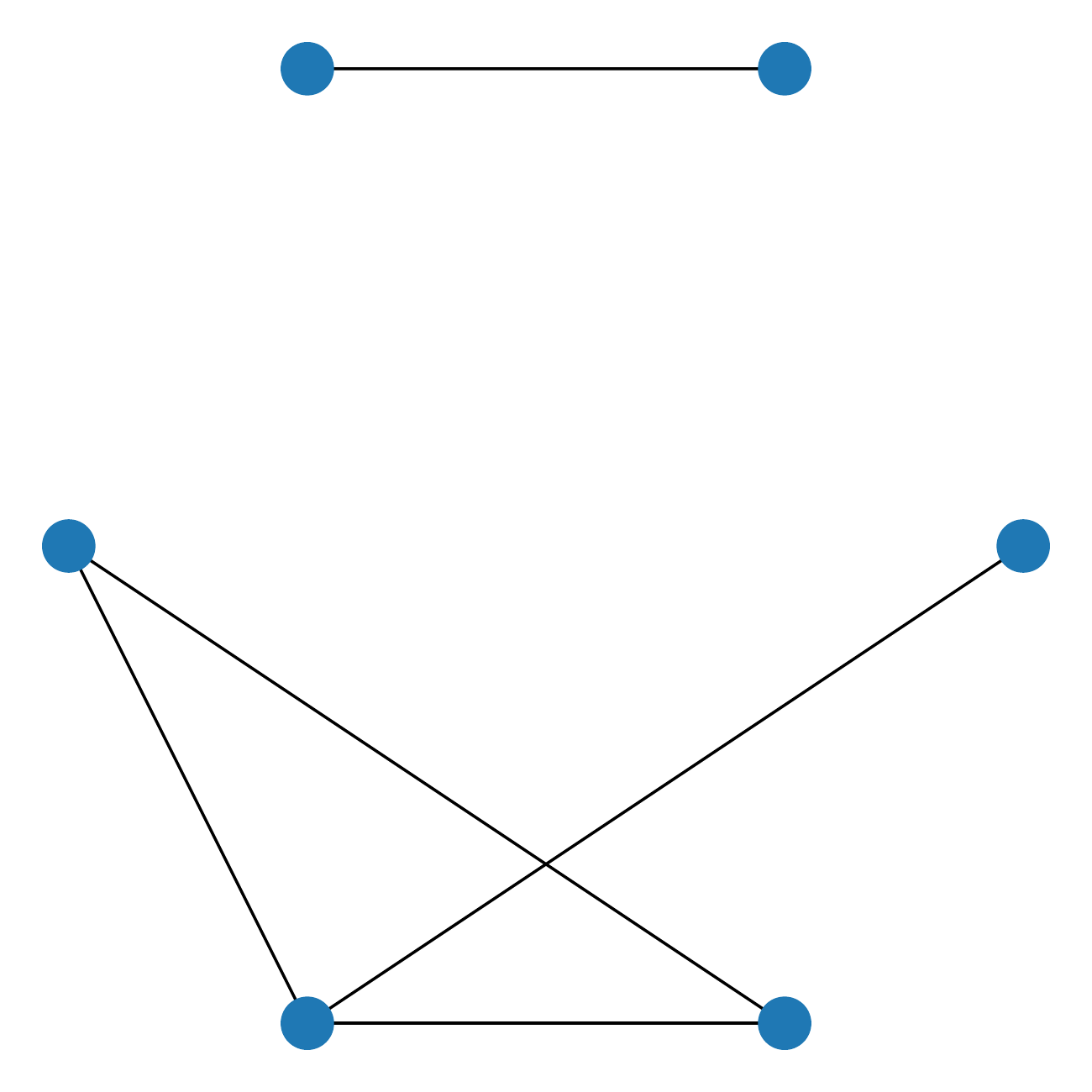}
        \caption{$s=5$}
        \label{fig:5linegraph}
\end{subfigure}
\vspace{0.5em}
\end{adjustbox}
  \caption{\small \textit{Line graphs computed from the virology genomics data \cite{FeSHeE20}. They are plotted using NetworkX in Shell layout. The six most important genes in the original hypergraph are identified by the $5$-line graph, which are ISG15, IL6, AFT3, RSAD2, USP18 and IFIT1. }}  \label{fig:ensemble_linegraphs}

\vspace{-.2em}  
\end{figure}

\section{Real-world Applications}
\label{sec:applications}

In this section, we illustrate the utility of our framework using three real-world applications: identifying the most important genes in a transcriptomics data, revealing strong co-authorships among authors, and uncovering collaboration networks among actors on Internet Movie Database (IMDB).

\subsection{Identifying Genes Critical to Pathogenic Viral Response}
\label{sec:gene-exam}

Though graph models are quite successful in biological data modeling, they have limitations in representing complex relationships amongst entities. In biology, hypergraphs can be used to model gene and protein interaction networks.
Here we construct a hypergraph from the virology genomics data \cite{FeSHeE20}, where there are 9760 hyperedges representing genes, and 201 vertices representing individual biological samples with specific experimental ``conditions'' (e.g., mouse lung cells treated with a strain of Influenza virus and sampled at 8 hours). 
We omit the details of extracting the hypergraphs from the dataset due to space constraint.

To identify important genes in this hypergraph, we compute the $s$-connected components and the $s$-betweenness centrality scores of the vertices within each $s$-connected component. \Cref{fig:ensemble_linegraphs} shows these $s$-line graphs. As the $s$ value is increased, the important  genes are clearly identifiable in the visualization. In particular, gene IFIT1 and USP18 have the highest centrality scores, implying that they are the two most important genes. They share more than 100 vertices between them. This indicates that IFIT1 and USP18 are both perturbed in over 100 experimental conditions at the same time. Our $s$-line graphs clearly reveal the strength of the connections of those two genes that previous graph-based models are unable to deduce.

\subsection{Revealing Relationships Among Authors}\label{sec:author-exam}

For certain hypergraph analytics, the formation of an ensemble of $s$-line graphs is strictly necessary. To illustrate a particular type of analysis that necessitates $s$-line graph construction, 
we construct a hypergraph from the condensed matter author-paper network in Los Alamos e-Print Archive~\cite{structure_newman_2001}. This hypergraph contains 16,726 authors as vertices, 22,016 papers as hyperedges, and 58,595 author-paper inclusions.

To reveal the relationships among authors in this network, we compute an ensemble of $s$-line graphs where $s$ ranges from 1 to 16 (16 is the max $s$ that produces non-singleton components). We compute the normalized algebraic connectivity of the $s$-line graphs of author-paper dataset. Normalized algebraic connectivity is the second-smallest eigenvalue of the normalized Laplacian matrix\cite{Fiedler1973,chung1997spectral}; larger values imply stronger connectivity properties of the $s$-line graph and hence the hypergraph.

\begin{wrapfigure}{l}{.28\textwidth}
\vspace{-1.4em}
\begin{adjustbox}{minipage=\columnwidth,scale=1.85}
 \includegraphics[width=0.30\textwidth]{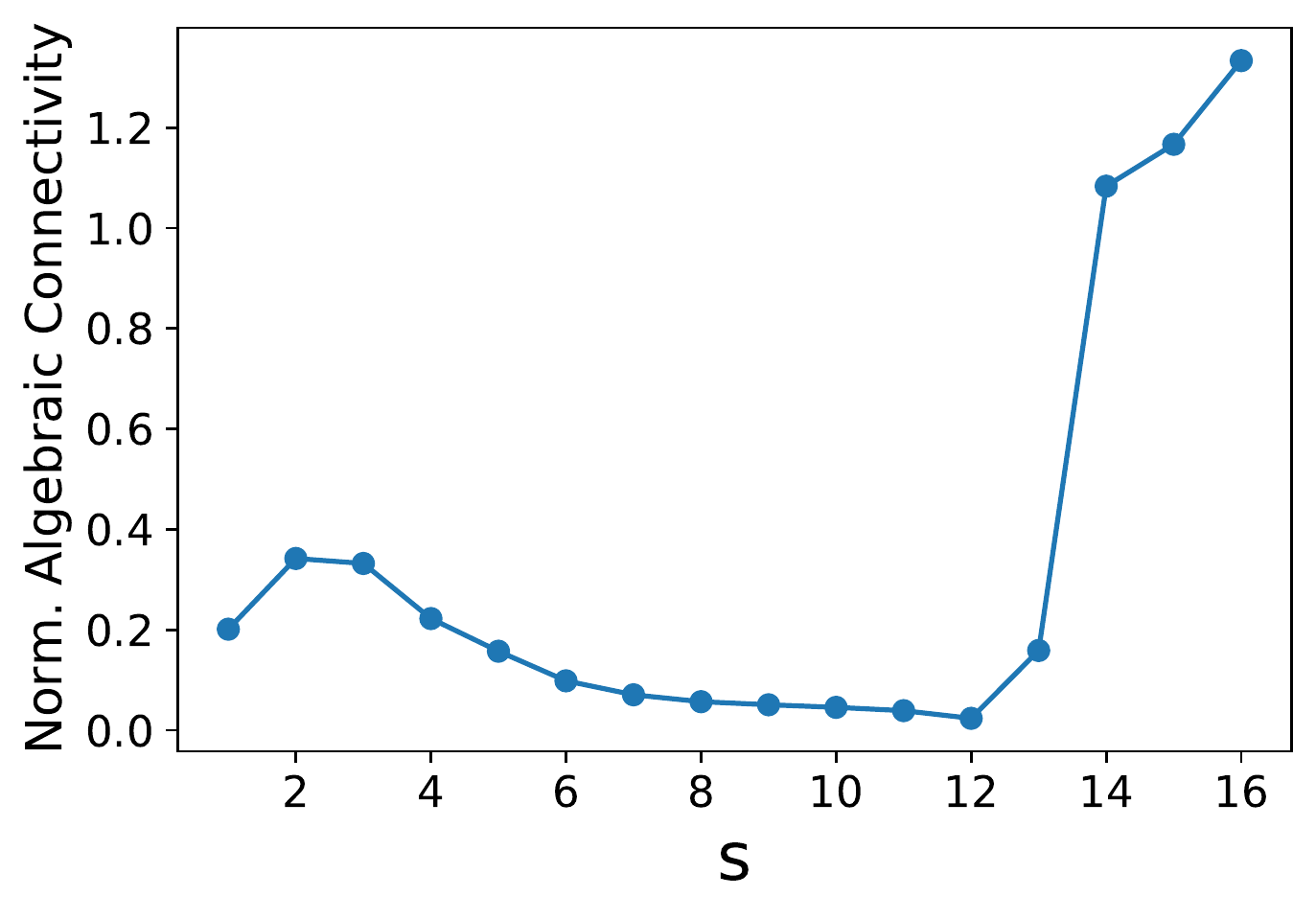}
 \end{adjustbox}
    \caption{\small \textit {Normalized algebraic connectivity for condensed matter author-paper 
    network.
    } }\label{fig:algebraic_condmat}
\vspace{-1.23em}
\end{wrapfigure}
As observed from \Cref{fig:algebraic_condmat}, decreasing values of algebraic connectivity from $s$=3 to $s$=12 reveals that many authors collaborate on papers only sparsely, meaning the vertices (authors) within a  connected component are sparsely connected with each other. However, the sharp increase in algebraic connectivity starting from $s$=13 demonstrate the fact that authors who have co-authored at least in 13 papers are more likely to collaborate with each other (signified by the denser connections within a connected component of an $s$-line graph). 
In this way, eigenvalues can provide insight into how well each of the connected components in an $s$-line graph remains connected and consequently provide insight about the original hypergraph connectivity. In addition, as the $s$ value grows, these techniques can assist in understanding how well the connectivity is preserved. 

\subsection{Uncovering Collaborations Among Actors}

Consider uncovering groupings of actors who have collaborated on at least $s$ movies. We can query this information from 
Internet Movie Database (IMDB) by constructing a hypergraph (where the movies are vertices, and actors are hyperedges), and computing the $s$-line graphs.
We compute $s$-connected components and $s$-betweenness centrality on these $s$-line graphs.
We start by working on three database tables from the database: \texttt{title.basic}, \texttt{name.basic} and \texttt{title.principals}~\cite{imdb_interfaces}. These tables contain approx. 11 million titles, approx. 8 million actor names, and approx. 18 million principal cast/crew for titles respectively.

The three collaboration networks that we uncovered within IMDB are reported below. 
Only the actors having a non-zero centrality scores are shown.
These actors collaborated in more than 100 movies together:
{
\footnotesize
\vspace{-0.2em}
\begin{minted}[escapeinside=||,gobble=0]{python}
(compute s-connected components) 4 us
Here are the 100-connected components:
[Adoor Bhasi, Bahadur, Paravoor Bharathan, Jayabharati, 
Prem Nazir], [Matsunosuke Onoe, Suminojo],
[Kijaku Ôtani, Kitsuraku Arashi],[Panchito, Dolphy]. 
(compute s-betweenness centrality) 15 us
Adoor Bhasi(0.1111), Matsunosuke Onoe(0.0111), 
Kijaku Ôtani(0.0111) //normalized score
\end{minted}  
\vspace{-0.2em}
}
We observe that, for the network in which Adoor Bhasi is a member, he has a centrality score of 0.11, while others have a score of 0. This means that Adoor Bhasi is the most important actor. 
Specifically, this network is a star graph where Adoor is the center vertex because all the other actors have a zero centrality score. 
Previous multigraph-formulation approach implemented in Python to compute betweenness centrality along took 10 hours on a Windows 10 machine (a 3.2 GHz CPU with 8 GB RAM)~\cite{lewis2020centre}. On the other hand,
our implementation took a total of 80ms to execute on a Mac Mini (M1 chip, with 16GB RAM) to compute the 100-line graph, 100-connected components and 100-betweenness centrality. 



\section{Experimental Analysis}
\label{sec:expr}
In this section, we evaluate the performance of our $s$-line graph algorithms in comparison with the algorithms proposed in~\cite{firoz_2020_efficient} and an efficient SpGEMM algorithm. We also discuss the scalability, workload characteristics and evaluation of the workload balancing techniques of our proposed algorithms.  \Cref{table:algo_notation} summarizes the shorthand notations we use for different algorithms with different workload distribution strategies.

\subsection{Experimental Setup}
Our experiments are run on a machine with a two-socket Intel Xeon Gold 6230 processor, having 20 physical cores per socket, each running at 2.1 GHz, and 28 MB L3 cache. The system has 188 GB of main memory. Our code is implemented in \texttt{C++17}, parallelized with 
Intel oneTBB 2020.3, and compiled with GCC 10.2 compiler and \texttt{-Ofast -march=native} compilation flags.

\subsection{Dataset}

We conducted experiments with real-world hypergraphs (Table~\ref{tab:input_hypergraph_prop}) from various domains, ranging from social to cyber to web.
The activeDNS (ADNS) dataset from Georgia Institute of Technology contains mappings from domains to IP addresses~\cite{ActiveDNSdataset}. When constructing hypergraphs with ADNS dataset, we consider the domains as the hyperedges and IPs as vertices. 
Additionally, we ran our experiments with datasets curated in~\cite{shun2020practical}. For these curated datasets, in particular, each hypergraph, constructed from the social network datasets such as com-Orkut and Friendster in \Cref{tab:input_hypergraph_prop}, are materialized by running a community detection algorithm on the original dataset obtained from Stanford Large Network Dataset Collection (SNAP)~\cite{leskovec2016snap}. In the resultant hypergraphs, each community is considered as a hyperedge and each member of a community as a vertex. Other larger datasets include 
Web, and LiveJournal, collected from Koblenz Network Collection (KONECT)~\cite{kunegis2013konect} as bipartite graphs. 

Additionally, we selected two large datasets: Amazon-reviews~\cite{amazon_data} (where hyperedges are sets of product reviews on Amazon, and nodes are product categories) and Stackoverflow-answers~\cite{austin_data} (where hyperedges are sets of questions and nodes are the tags for questions answered by users on Stack Overflow).

\begin{table}[ht]
\small
\vspace{-0.5em}
\centering
\begin{tabular}{ccll}
\toprule
Notation       & Algo.  & Partitioning & Relabel-by-degree\\
\midrule
$1BA$ & Algo.~\underline{\ref{algo:efficient_s_overlap_serial}} & \underline{B}locked & \underline{A}scending \\
$1BD$ & Algo.~\underline{\ref{algo:efficient_s_overlap_serial}} & \underline{B}locked & \underline{D}escending \\
$1BN$ &  Algo.~\underline{\ref{algo:efficient_s_overlap_serial}} & \underline{B}locked & \underline{N}o \\
$1CA$ & Algo.~\underline{\ref{algo:efficient_s_overlap_serial}} & \underline{C}yclic & \underline{A}scending \\ 
$1CD$ & Algo.~\underline{\ref{algo:efficient_s_overlap_serial}} & \underline{C}yclic & \underline{D}escending \\
$1CN$ & Algo.~\underline{\ref{algo:efficient_s_overlap_serial}} & \underline{C}yclic & \underline{N}o \\
$2BA$ & Algo.~\underline{\ref{algo:map_s_overlap_serial}} & \underline{B}locked & \underline{A}scending \\
$2BD$ & Algo.~\underline{\ref{algo:map_s_overlap_serial}} & \underline{B}locked & \underline{D}escending \\
$2BN$ &  Algo.~\underline{\ref{algo:map_s_overlap_serial}} & \underline{B}locked & \underline{N}o \\
$2CA$ & Algo.~\underline{\ref{algo:map_s_overlap_serial}} & \underline{C}yclic & \underline{A}scending \\
$2CD$ & Algo.~\underline{\ref{algo:map_s_overlap_serial}} & \underline{C}yclic & \underline{D}escending \\
$2CN$ &  Algo.~\underline{\ref{algo:map_s_overlap_serial}} & \underline{C}yclic & \underline{N}o  \\
\bottomrule
\end{tabular}
\vspace{0.5em}
\caption{\small \textit{Notation for different algorithms with different partitioning techniques and relabel-by-degree ordering. }
}\label{table:algo_notation}
\vspace{-1.0em}
\end{table}

\setlength{\tabcolsep}{12pt}

\begin{table}[ht]
  \centering
  \small
  \tabcolsep=0.05cm
  \begin{tabular}{*{10}{c}} \toprule
    Type & hypergraph & $|V|$ & $|E|$ & $\overline{d}_v$ & $\overline{d}_e$ & $\Delta_{v}$ & $\Delta_{e}$   \\ \midrule
\multirow{3}{*}{Social} & com-Orkut  & 2.3M & 15.3M & 46 & 7 & 3k & 9.1k  \\
                             & Friendster  & 7.9M &1.6M & 3 & 14 & 1.7k & 9.3k   \\ 
                             & LiveJournal & 3.2M & 7.5M & 35 & 15 & 300 & 1.1M  \\\hline
\multirow{3}{*}{Web} & Web & 27.7M & 12.8M & 5 & 11 & 1.1M & 11.6M  \\
& Amazon-reviews & 2.3M &  4.3M & 32 &  17 & 29k & 9.4k \\
& Stackoverflow-answers & 1.1M &  15.2M & 2 & 24 &  356 &  61.3k \\\hline
\multirow{1}{*}{Cyber} & activeDNS & 4.5M & 43.9M & 11 & 1 & 714.6k & 1.3k \\ \hline
\multirow{1}{*}{Email} & email-EuAll & 265.2k & 265.2k & 2 & 2 & 7.6k & 930 \\
    \bottomrule
    \end{tabular}
    \vspace{0.5em}
    \caption{\small \textit{Input characteristics. The number of vertices ($|V|$) and hyperedges ($|E|$) along with the average degree ($\overline{d}$), and maximum degree ($\Delta$) for the hypergraph inputs are tabulated here.
    All the hypergraphs have a skewed hyperedge degree distribution. }}\label{tab:input_hypergraph_prop}
\vspace{-0em}
\end{table}

\begin{figure}[tp]
\vspace{-1em}
\centering
\begin{adjustbox}{minipage=\columnwidth,scale=0.85}
\includegraphics[width=\linewidth]{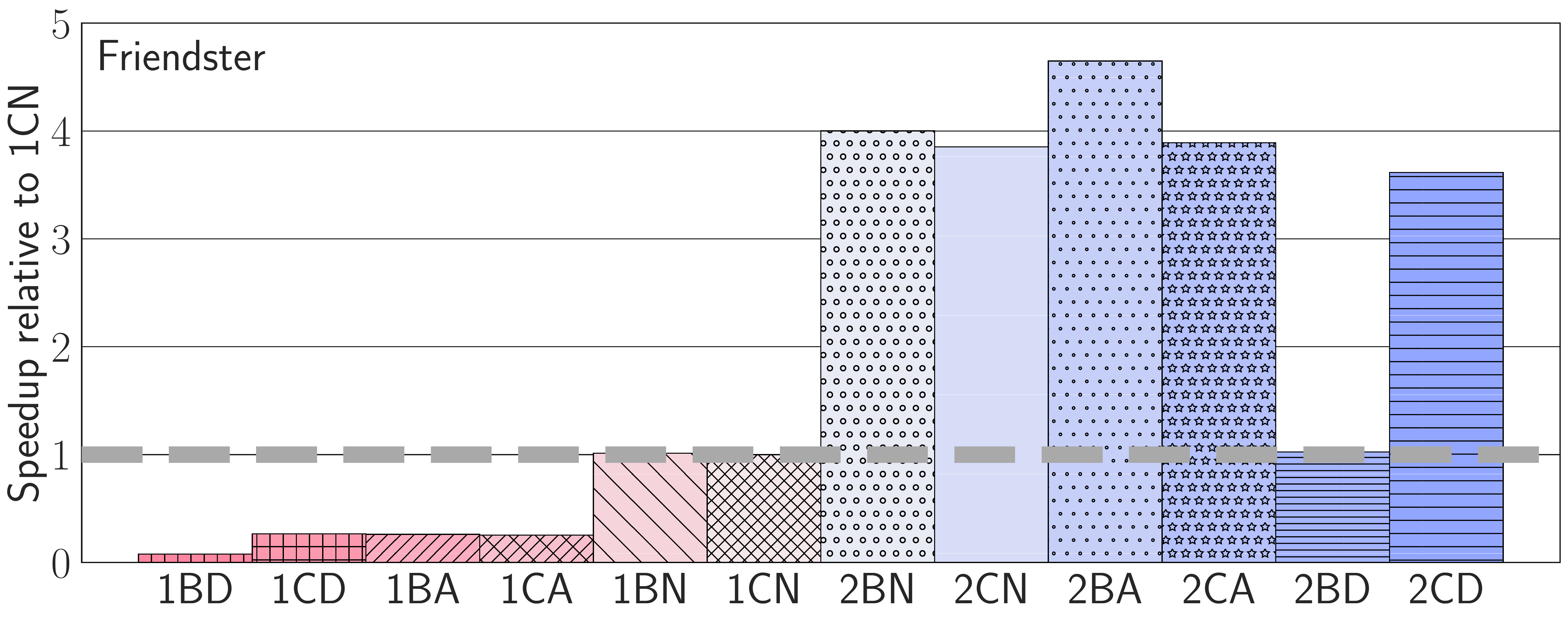}
\includegraphics[width=\linewidth]{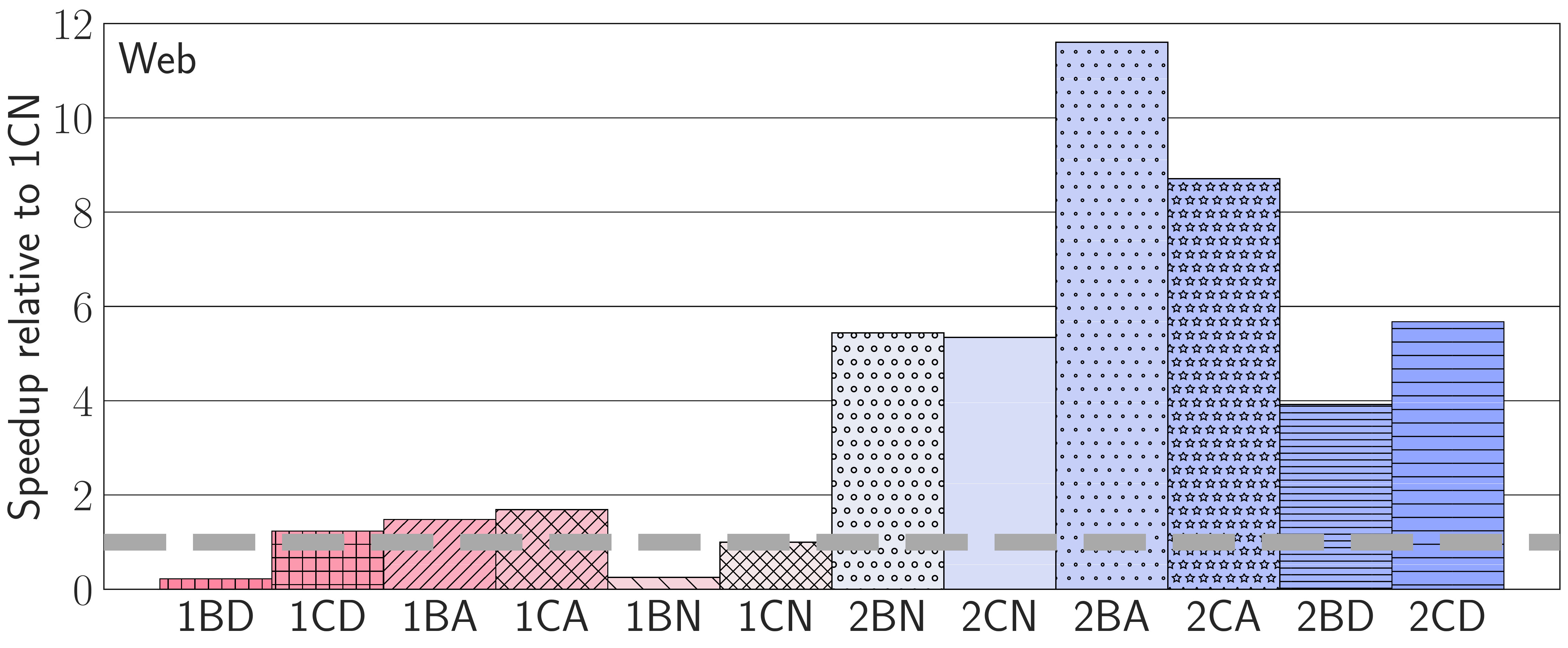}
\includegraphics[width=\linewidth]{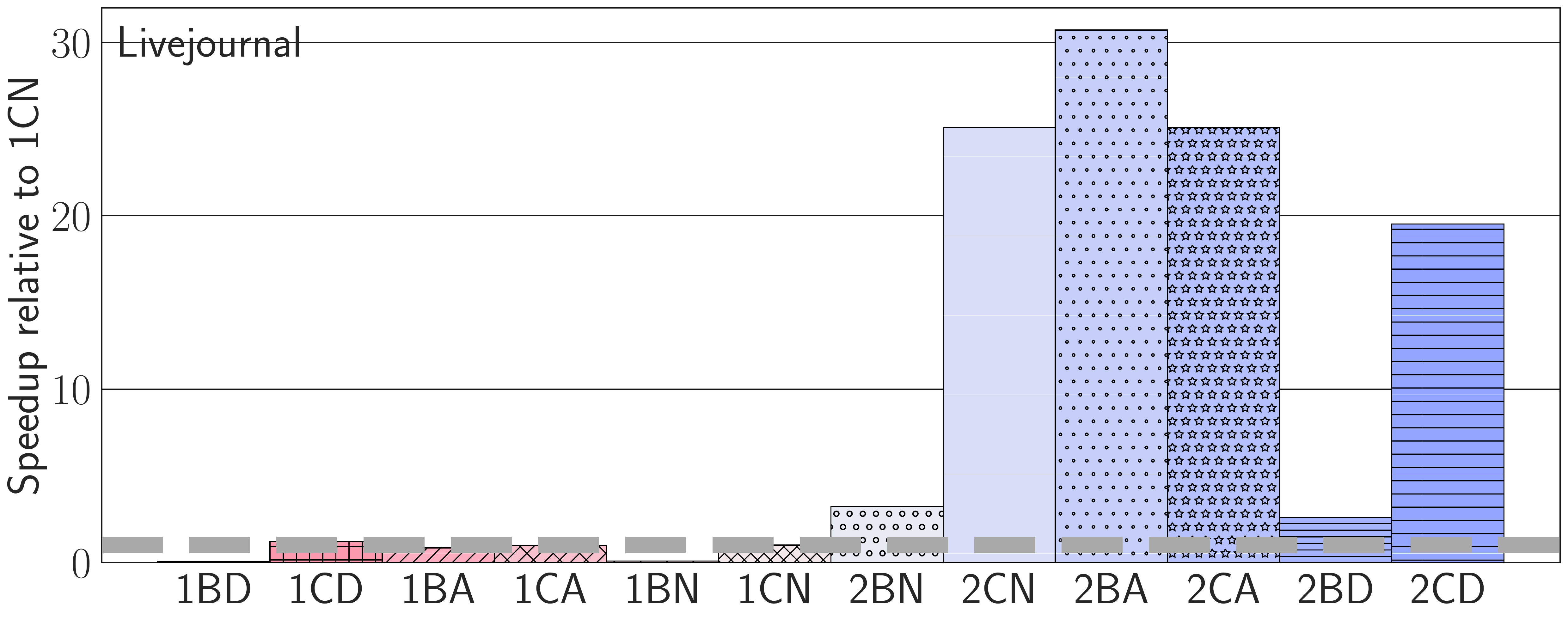}
\includegraphics[width=\linewidth]{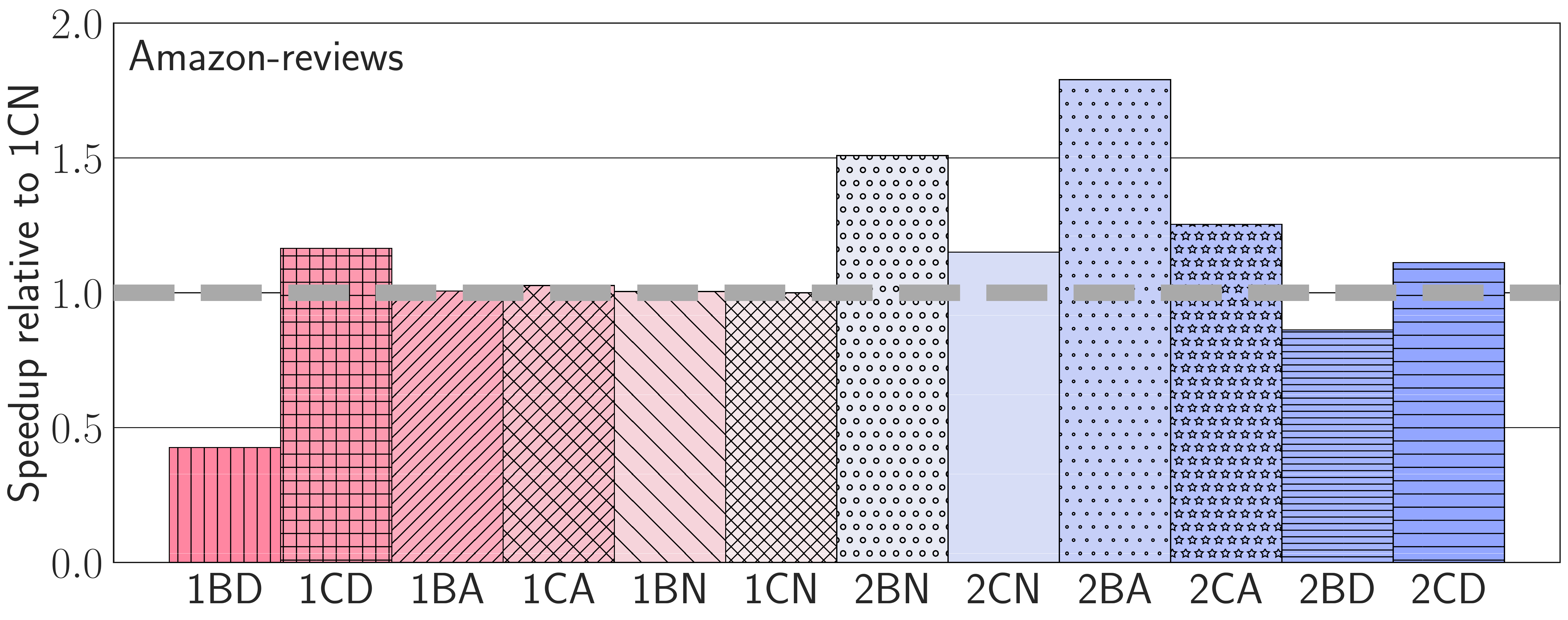}
\includegraphics[width=\linewidth]{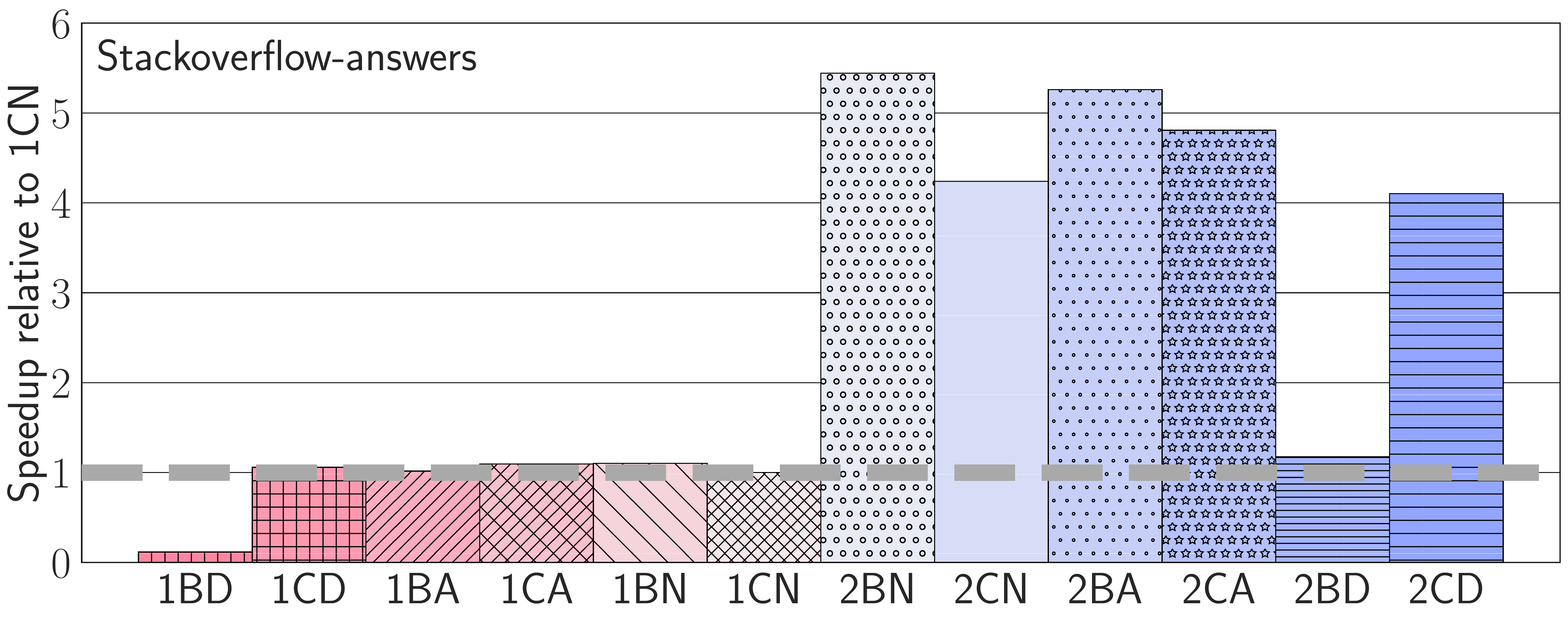}
\end{adjustbox}
\vspace{0.2em}
\caption{\small \textit{Speedup relative to \Cref{algo:efficient_s_overlap_serial} with cyclic work distribution (1CN) where $s=8$. 
}}\label{fig:eval_algo} 
\end{figure}

\vspace{-0.4em}

\subsection{Performance Analysis}

In~\Cref{fig:eval_algo}, we report the performance of different algorithms listed in~\Cref{table:algo_notation}. The execution time for each algorithm is normalized w.r.t. 1CN (Algorithm~\underline{\ref{algo:efficient_s_overlap_serial}} with \underline{c}yclic distribution and \underline{n}o relabeling). 
Here, we do not report results of~\Cref{algo:map_s_overlap_ensemble_serial}, as it
fails on most of the datasets (except for email-EuAll) due to its memory limitation.

As observed from~\Cref{fig:eval_algo}, our algorithm (\Cref{algo:map_s_overlap_serial}), in conjunction with the right combination of workload distribution strategy and relabel-by-degree, performs best and achieves ${\approx} 5{\times}{-}31{\times}$ speedup for Web, and LiveJournal datasets. Larger inputs with skewed degree distribution (containing a handful of high-degree hyperedges) perform best when run with 2BA (Algorithm~\underline{\ref{algo:map_s_overlap_serial}} with \underline{b}locked distribution and hyperedges relabeled by degrees in \underline{a}scending order). 
Interestingly, relabeling the edges based on their degrees (both ascending and Descending) does not provide drastic performance benefit for Friendster, Amazon-reviews and Stackoverflow-answers. 
These 3 datasets have 
smaller maximum degrees ($\Delta_{e}$). Hence, relabel-by-degree does not provide significant benefit in improving the performance. In this case, the additional overhead of relabeling the hyperedges based on degrees heavily penalizes the execution time (we included the pre-processing time to relabel by degree in the total execution time). 


\subsection{Strong Scaling}

We conducted strong scaling experiments for our algorithms with different hypergraph inputs and we report the results in~\Cref{fig:strong_scaling}. Here we double the number of threads while keeping the input size constant. The performance of the algorithms improves up to 16 threads. Beyond 16 threads, performance does not improve significantly. For inputs with highly-skewed degree distribution (LiveJournal, com-Orkut, Web), 
2CA demonstrates best scaling behaviour, as cyclic distribution enables better load balancing. Both block and cyclic distributions without relabeling achieve similar performance. 

\begin{figure}
\vspace{-1.5em}
  \centering
\includegraphics[width=.48\linewidth]{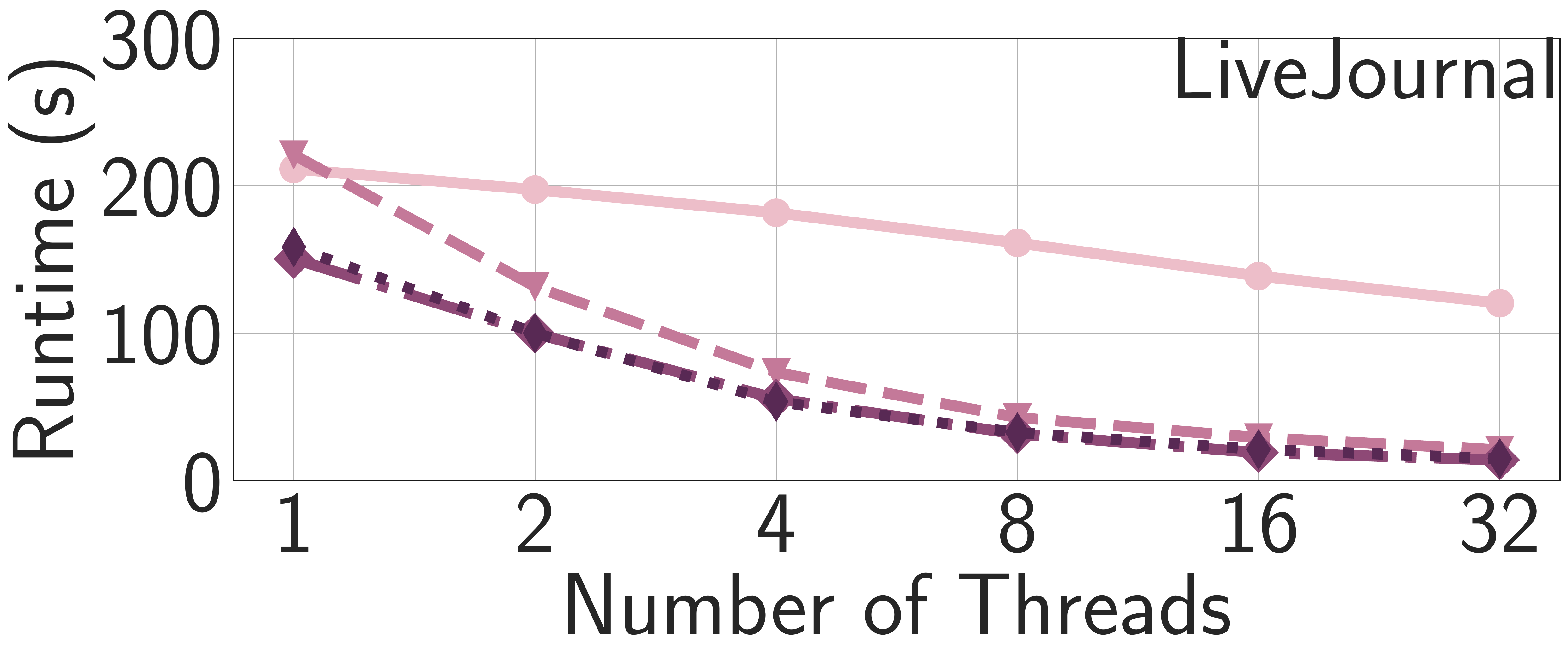}
\includegraphics[width=.48\linewidth]{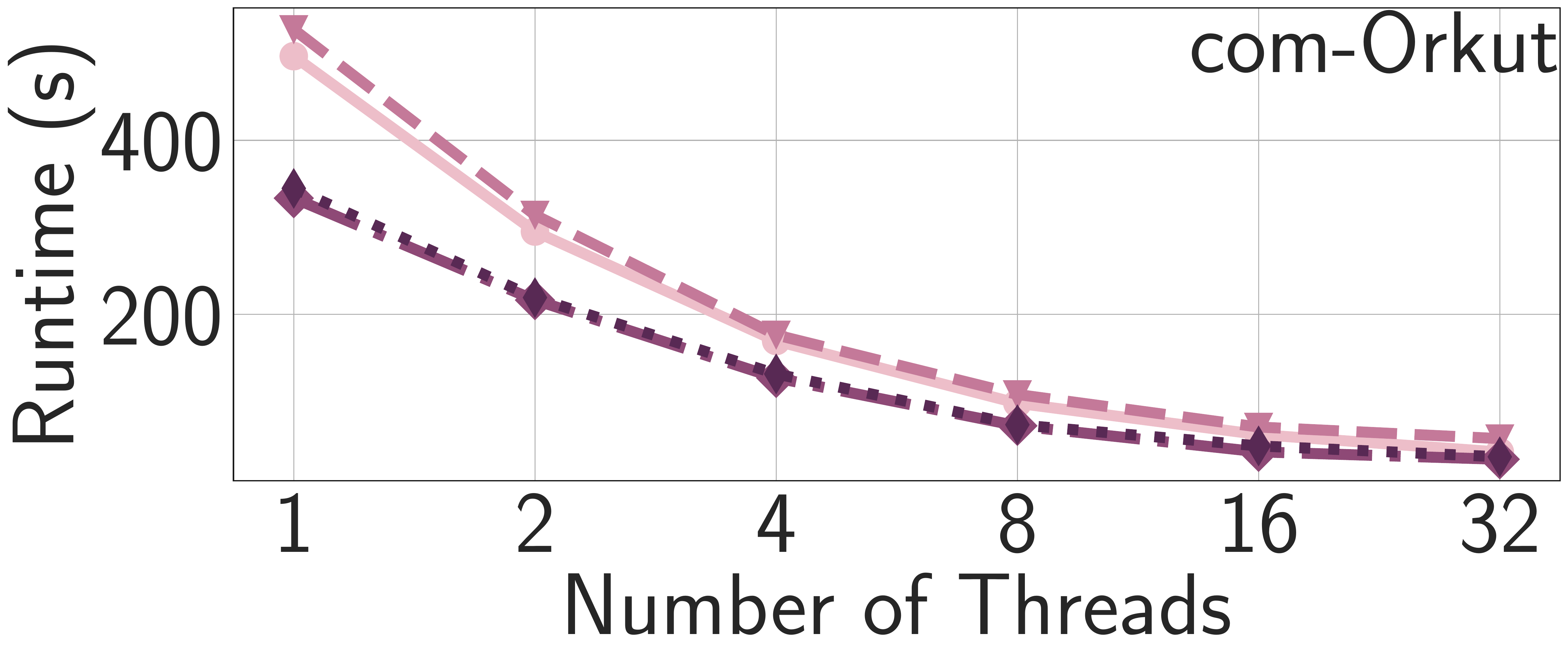}
\includegraphics[width=.48\linewidth]{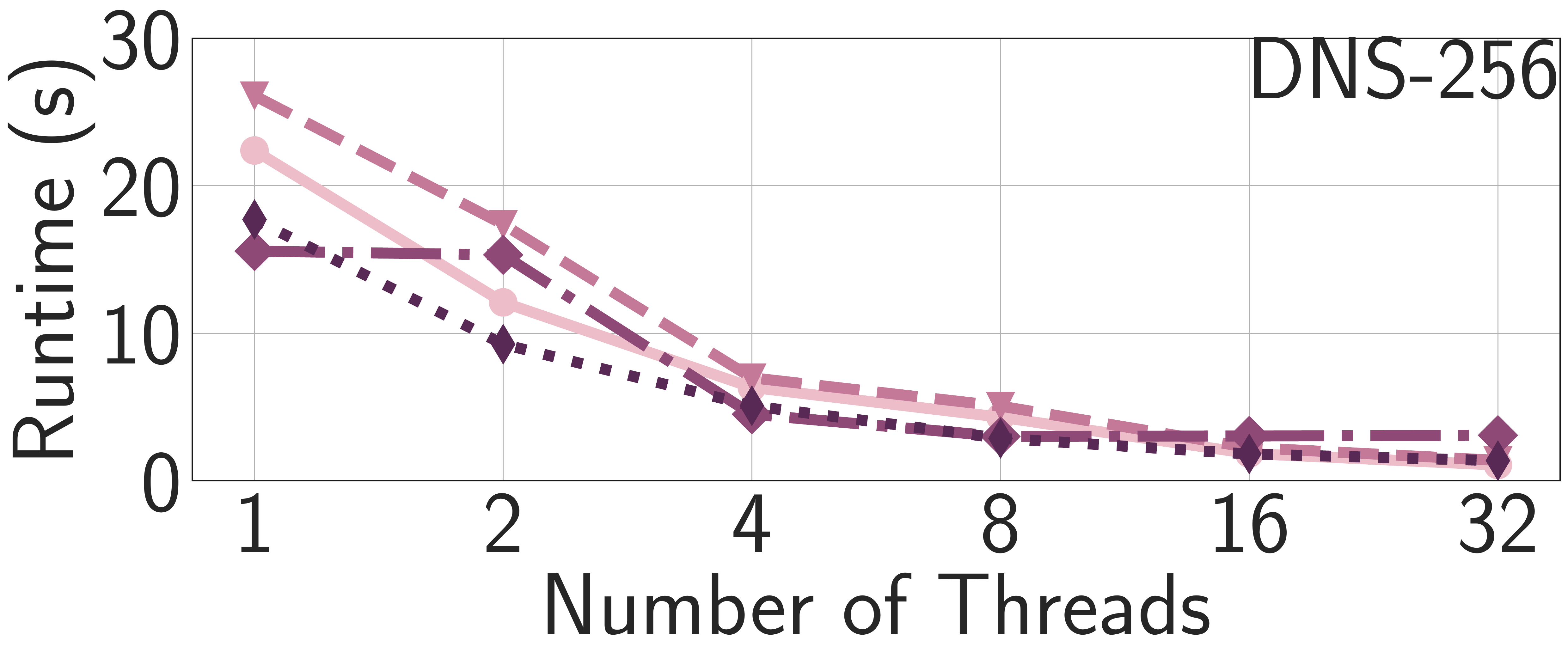}
\includegraphics[width=.48\linewidth]{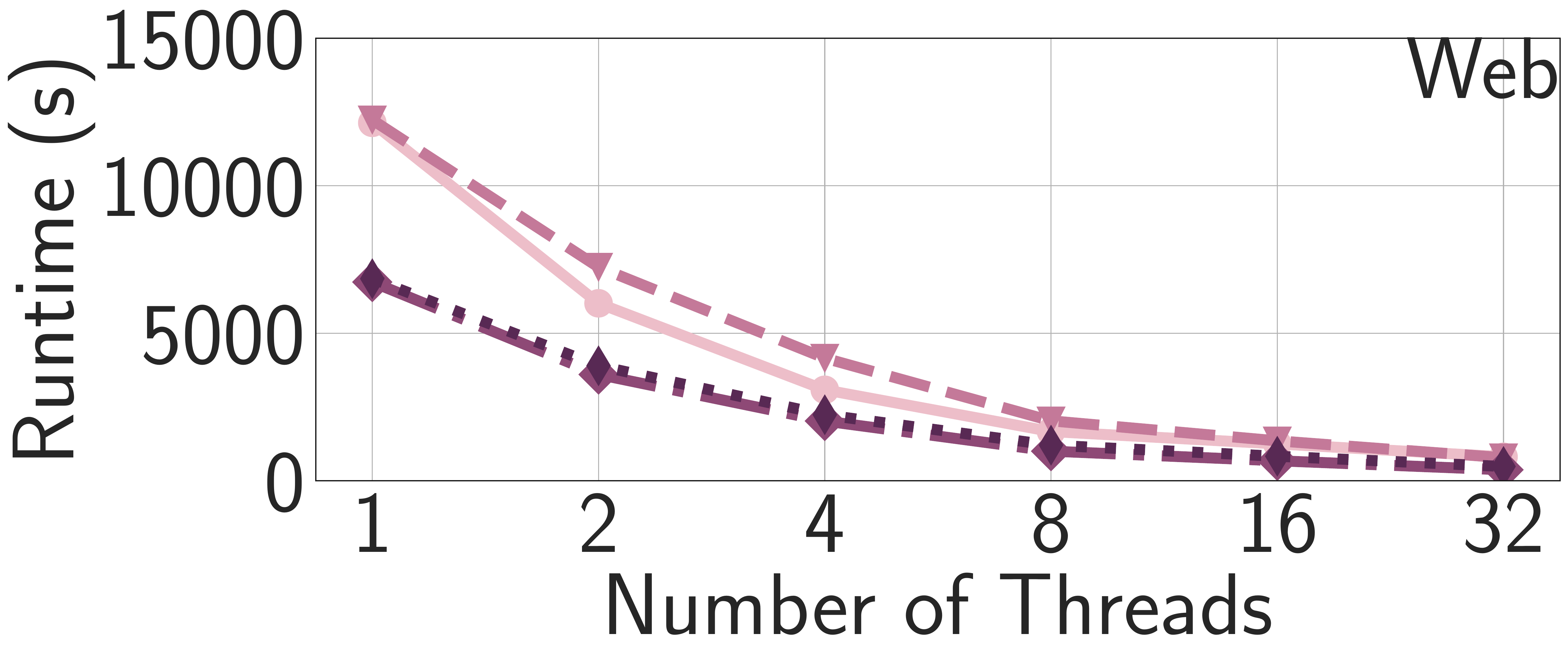}
\includegraphics[width=.6\linewidth]{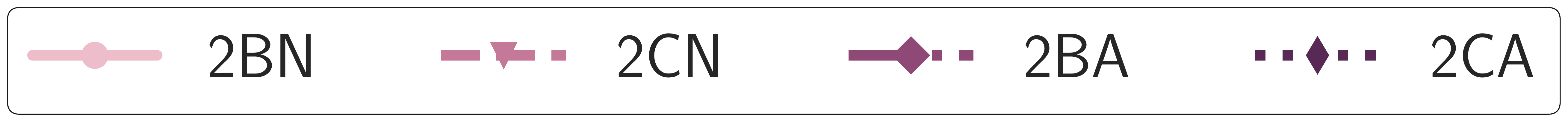}
\vspace{0.3em}
\caption{\small \textit{Strong scaling results with blocked distribution and cyclic distribution for \Cref{algo:map_s_overlap_serial} when $s=8$. 
}}\label{fig:strong_scaling} 
\end{figure}

\subsection{Weak Scaling}

\begin{figure}
\centering
\vspace{-.5em}
  \includegraphics[width=0.42\textwidth, ]{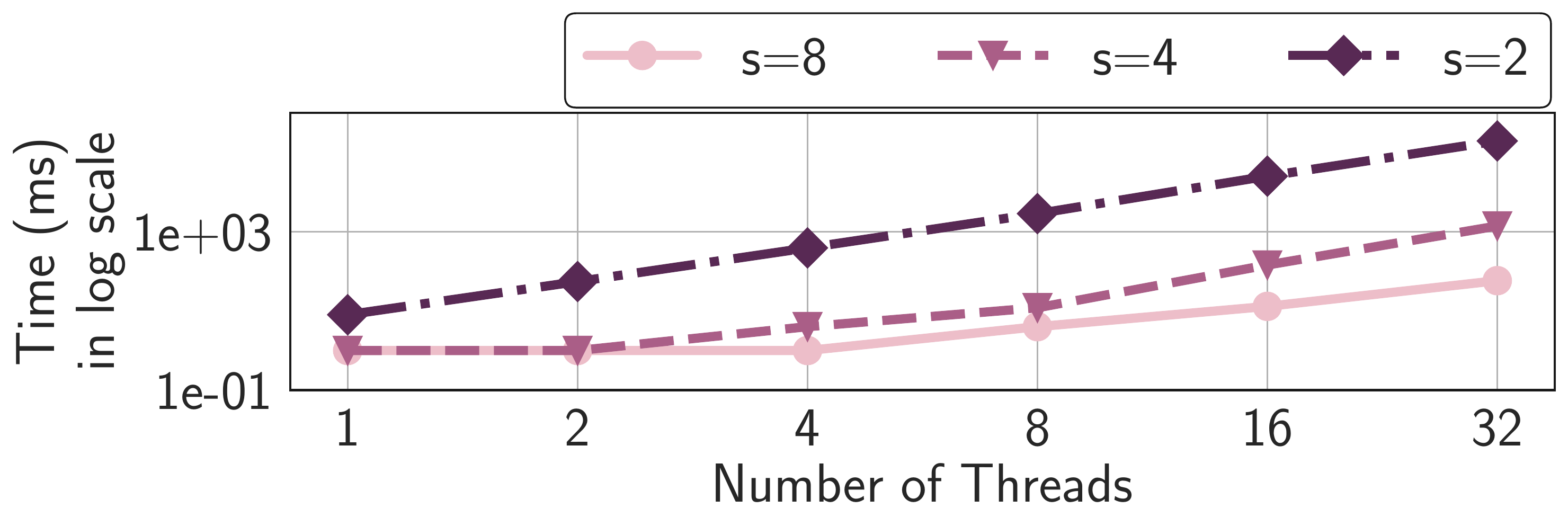}
  \label{fig:dns_full_blocked}
\caption{\small \textit{Weak scaling results of \Cref{algo:map_s_overlap_serial} using blocked workload distribution for activeDNS dataset. 
} }
  \label{fig:weak_scaling_hypergraph}
 \vspace{-.5em}
\end{figure} 
We performed weak scaling experiments of \Cref{algo:map_s_overlap_serial} with the activeDNS dataset using blocked workload distribution strategy. Here we approximately double the size of the hypergraph (workload) as we double the number of threads (computing resources). We start with 4 AVRO files worth of data (dns\_4) and scale up to 128 files (dns\_128). 
With larger $s$ values, the performance of the algorithms improves (\Cref{fig:weak_scaling_hypergraph}).

\subsection{Workload Characterization}
\Cref{fig:workload-dist-livejournal} shows the number of hyperedges visited by each thread in the innermost loop of \Cref{algo:map_s_overlap_serial} with different partitioning strategies for LiveJournal dataset.  
\begin{figure}
\centering
    {\includegraphics[width=.42\textwidth]{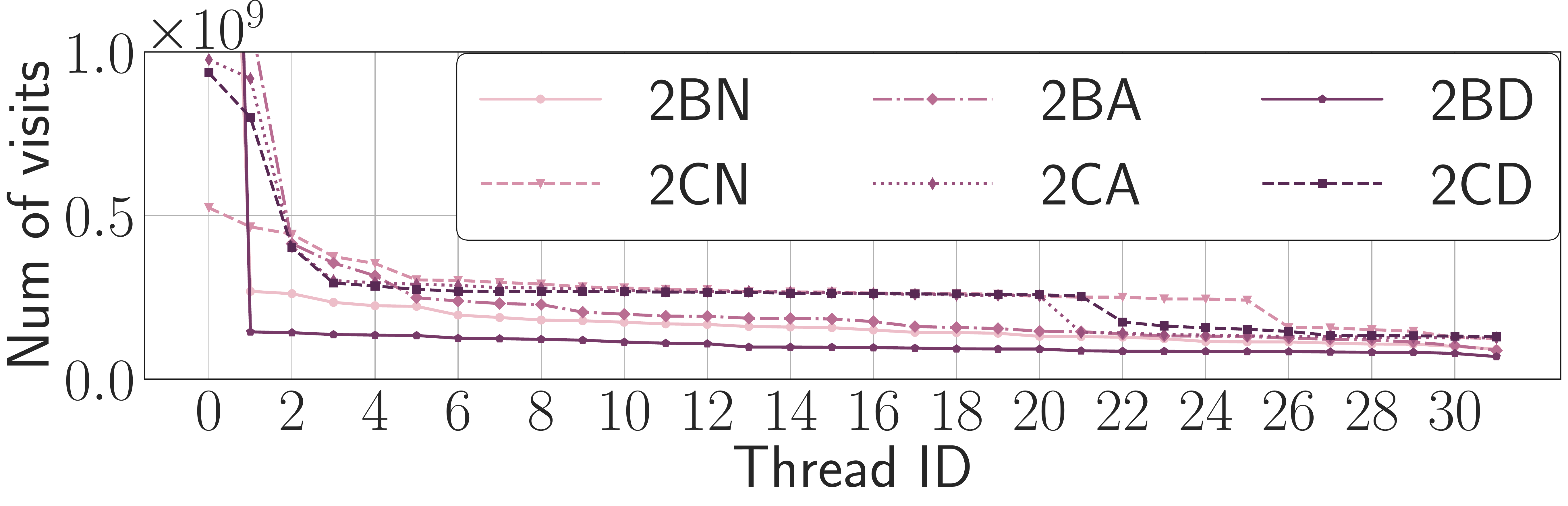}}
    \caption{\small \textit{Workload distribution among 32 threads when partitioning the hyperedges (outermost loop of the s-line graph algorithms) in a blocked or cyclic manner in \Cref{algo:map_s_overlap_serial} for LiveJournal input.}} 
  \label{fig:workload-dist-livejournal}
\vspace{0em}
\end{figure}  
As can be observed from ~\Cref{fig:workload-dist-livejournal}, without relabel-by-degree, cyclic distribution achieves better workload balance than blocked distribution.
We also observe in ~\Cref{fig:eval_algo} that blocked or cyclic distribution, in conjunction with relabeling by degree in ascending order, performs best overall. We investigated this observation in details with Intel VTune Profiler and found out that relabel-by-degree in ascending order provides more favorable cache reuse (due to almost 0.5x less LLC cache misses) to \Cref{algo:map_s_overlap_serial} than the descending order.

\subsection{Comparison with an SpGEMM-based Approach}
\label{sec:spgemm-cmp}
\begin{figure}[ht]
\vspace{-0.8em}
\centering
  \includegraphics[width=.42\textwidth]{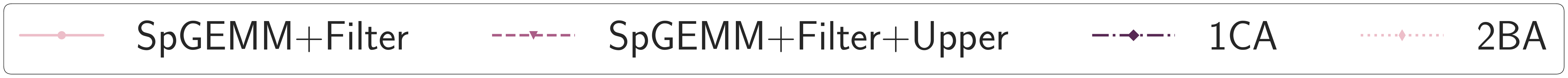}\\
  \includegraphics[width=.42\textwidth]{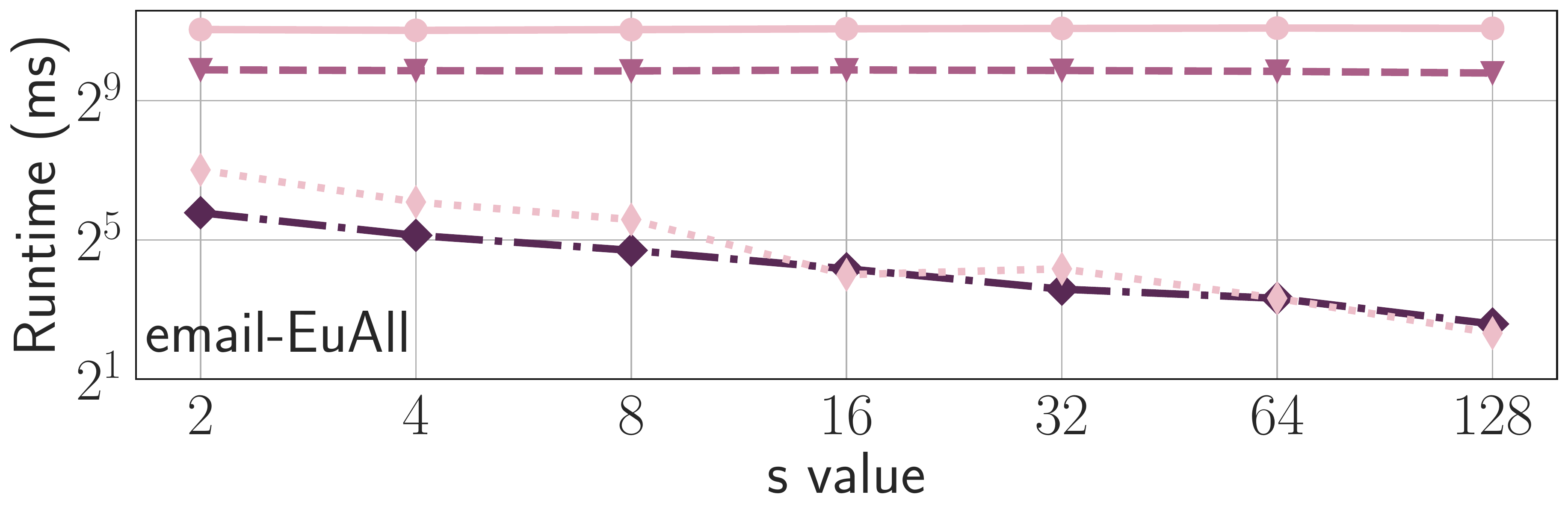}
  \includegraphics[width=.42\textwidth]{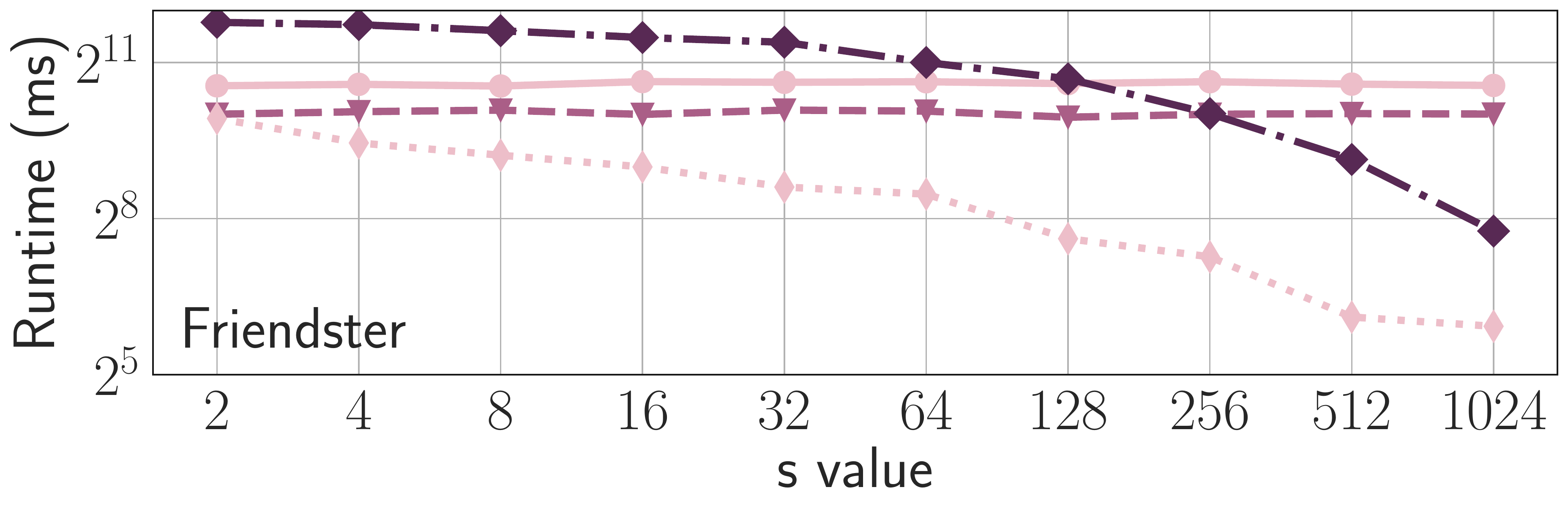}
  \caption{\small \textit{Comparison of \Cref{algo:efficient_s_overlap_serial}, and  \Cref{algo:map_s_overlap_serial} with an SpGEMM-based approach. Here SpGEMM+Filter+Upper refers to only consider the upper triangular part of the adjacency matrix.}}
  \label{fig:spgemm_map}
\vspace{-.5em}  
\end{figure}

We also compare the performance of our hashmap-based algorithms and ~\Cref{algo:efficient_s_overlap_serial} with a state-of-the-art SpGEMM-based library~\cite{SpGEMMrepo}. We modified the SpGEMM code to add the filtration step, and to only consider the upper triangular part of the matrix. Here, the SpGEMM library first computes $HH^T$, and then filters the edges with at least $s$ overlaps. We report the results with email-EuAll and Friendster datasets. The SpGEMM library fails to run on other larger hypergraph datasets. 
The results are reported in~\Cref{fig:spgemm_map}. 
With all datasets and for different $s$ values, \Cref{algo:map_s_overlap_serial} runs faster than the SpGEMM+Filter+Upper algorithm. The efficient algorithm  (\Cref{algo:efficient_s_overlap_serial}) runs faster than the SpGEMM+Filter+Upper algorithm with the email-EuAll 
dataset, but slower than the SpGEMM+Filter+Upper algorithm with Friendster dataset (for smaller $s$ values).
With larger $s$ 
values 
in all cases, our algorithm is orders of magnitude faster than the SpGEMM+Filter+Upper approach. 
The improvement can be attributed to the degree-based pruning.
Note that computation of the $s$-line graphs with higher $s$ values ($s=1024$ for Friendster here) is still relevant, because, even with such a large $s$ overlap constraint, we found 20 connected components in the constructed $s$-line graph. This reveals that these 20 communities which share at least 1024 common members are the core of Friendster dataset.

Both the efficient and our hashmap-based algorithm are more suitable than off-the-shelf SpGEMM algorithm for the $s$-line graph computation. The SpGEMM algorithm is too general since it has to compute and store the product matrix before applying filtration upon the matrix. In contrast, our algorithm performs an in-place filtration. 
In addition, the SpGEMM+Upper algorithm performs half of the total work by only considering the upper triangular part of the hyperedge adjacency matrix. However, it is still orders of magnitude slower than our algorithm (especially with larger $s$ values).

\subsection{Comparison with the Clique-expansion Approach}
\label{sec:clique-expansion-cmp}


In~\Cref{tab:s1vs8}, we report the performance results of \Cref{algo:map_s_overlap_serial} when $s$=1 (the clique expansion graph) and $s$=8 on larger datasets. 
We ran the Label Propagation-based Connected Components (LPCC) after computing the $s$-line graphs with  \Cref{algo:map_s_overlap_serial} (2CA). 
With $s$=1, only Friendster and Livejournal datasets completed execution on a 128GB-memory machine.

\begin{table}[hb]
\vspace{-1em}
\small
\centering
{\begin{tabular}{p{0.05\columnwidth}|p{0.1\columnwidth}p{0.1\columnwidth}ll}
\hline
   &  Friendster   & LiveJournal   & com-Orkut  &  Web \\
\hline
s=1 & 12s & 76s & OOM &  OOM  \\
s=8 & 4s  & 31s & 59s  &  1510s \\
\hline
\end{tabular}}
\vspace{0.5em}
\caption{\small \textit{Execution time 
of $s$=1 (clique expansion)-based and $s$-line graph-based with $s$=8
Label-Propagation Connected Components (LPCC) with \Cref{algo:map_s_overlap_serial} 
(2CA). 
With $s$=1, com-Orkut and Web ran out of memory on a 128GB machine. The reported time includes end-to-end execution time of our framework. 
} \vspace{-1em} }\label{tab:s1vs8}

\end{table}

\vspace{-0.4em}
\section{Related Work}
\label{sec:relatedwork}

\rem{

For further background and discussion of the concepts of graphs, bipartite graphs and hypergraphs, we refer the reader to the books by Berge~\cite{berge1984hypergraphs,berge1973graphs,berge_hypergraphs_1989} and Bretto~\cite{bretto_hypergraph_2013} for hypergraph theory; Bondy and Murty~\cite{  BondyJ.A2008GT} and Pemmaraju and Skiena ~\cite{pemmaraju_computational_2003} for graph theory. We review in this brief section related works focusing primarily on algorithms.

Since then, a plethora of theoretical work has been published in the context of hypergraphs~\cite{chung1993laplacian,cooper2012spectra,katona1975extremal,dorfler1980category}. 

}

Hypergraph methods are well known for their applications in computer science;
for example, hypergraph partitioning enables parallel matrix
computations \cite{Devine2006} and
application in VLSI \cite{Karypis2000}.  In the
network science literature, researchers have devised several path and
motif-based hypergraph data analytics measures such as clustering coefficients
and
centrality metrics \cite{Estrada2006}.
Although an expanding body of research attests to the utility of hypergraph-based analyses \cite{battiston2020networks,iacopini_simplicial_2019,PaAPeG17}, and we are seeing increasingly wide adoption \cite{javidian_hypergraph_2020,LaNReJ20,MiM00}, 
many network science methods have been historically developed explicitly for graph-based analyses. 
Naik~\cite{naik2018intersection} wrote a survey on theoretical developments on line graphs. Bermond {\it et al.}\ \cite{bermond1977line} studied the properties of the $s$-line graphs of  hypergraphs. 

Shared-memory C++-based framework Hygra~\cite{shun2020practical}, and distributed-memory frameworks such as  Chapel-based CHGL~\cite{Jenkins_2018_chapel}, Apache Spark-based MESH \cite{heintz_mesh_2019} and HyperX \cite{jiang_hyperx_2019} presented a collection of efficient parallel algorithms for hypergraphs in their frameworks. These frameworks either rely on the original hypergraph or the expansion graphs of hypergraphs. None of the works except CHGL computes $s$-line graphs with $s>1$ and therefore cannot compute the $s$-walk measures. Moreover, in MESH/HyperX, on 8 compute nodes, a label-propagation-based connected component algorithm with clique expansion takes more than 2000s. In contrast, our framework takes $\approx$6s for the same computation, on a single-node.

\vspace{-0.4em}
\section{Conclusion}\label{sec:conclusion}
The notion of $s$-line graphs of a hypergraph is a novel way to  interpret relationships among different entities in a given dataset. In this paper, we have presented a scalable $s$-line graph computation framework by identifying a core set of stages required for end-to-end $s$-metric computation. We  proposed new parallel algorithms for $s$-line graph computations and explored different workload distribution strategies for our parallel algorithms in conjunction with considering relabel-by-degree and triangularization of the adjacency matrix as optimization techniques. We demonstrated that our algorithms outperform current state-of-the-art algorithms
. In particular,  hypergraphs with skewed-degree distribution can benefit from relabeling the hyperedge IDs by degrees. We showed that proper combination of algorithmic optimization and workload balancing technique can significantly improve the performance of the $s$-line graph computation stage, which is the most important and compute-intensive part of the framework. 
\vspace{-0.5em}

\section{Acknowledgement}\label{sec:ack}
This work was partially supported by the High Performance Data Analytics
(HPDA) program at the Department of Energy's Pacific Northwest National
Laboratory, and by the NSF awards IIS-1553528 and SI2-SSE 1716828. 
PNNL Information Release: PNNL-SA-167812. 
Pacific Northwest National Laboratory is operated by Battelle Memorial Institute under Contract DE-ACO6-76RL01830.

\bibliographystyle{IEEEtranS}
\bibliography{bibliography}

\begin{thebibliography}{10}
\providecommand{\url}[1]{#1}
\csname url@samestyle\endcsname
\providecommand{\newblock}{\relax}
\providecommand{\bibinfo}[2]{#2}
\providecommand{\BIBentrySTDinterwordspacing}{\spaceskip=0pt\relax}
\providecommand{\BIBentryALTinterwordstretchfactor}{4}
\providecommand{\BIBentryALTinterwordspacing}{\spaceskip=\fontdimen2\font plus
\BIBentryALTinterwordstretchfactor\fontdimen3\font minus
  \fontdimen4\font\relax}
\providecommand{\BIBforeignlanguage}[2]{{%
\expandafter\ifx\csname l@#1\endcsname\relax
\typeout{** WARNING: IEEEtranS.bst: No hyphenation pattern has been}%
\typeout{** loaded for the language `#1'. Using the pattern for}%
\typeout{** the default language instead.}%
\else
\language=\csname l@#1\endcsname
\fi
#2}}
\providecommand{\BIBdecl}{\relax}
\BIBdecl

\bibitem{aksoy_hypernetwork_2020}
S.~G. Aksoy, C.~Joslyn, C.~O. Marrero, B.~Praggastis, and E.~Purvine,
  ``Hypernetwork science via high-order hypergraph walks,'' \emph{EPJ Data
  Science}, vol.~9, no.~1, p.~16, 2020.

\bibitem{Barabasi2016}
A.-L. Barab{\'a}si, \emph{Network Science}.\hskip 1em plus 0.5em minus
  0.4em\relax Cambridge University Press, 2016.

\bibitem{battiston2020networks}
F.~Battiston, G.~Cencetti, I.~Iacopini, V.~Latora, M.~Lucas, A.~Patania, J.-G.
  Young, and G.~Petri, ``Networks beyond pairwise interactions: structure and
  dynamics,'' \emph{Physics Reports}, 2020.

\bibitem{berge1973graphs}
C.~Berge, \emph{Graphs and hypergraphs}.\hskip 1em plus 0.5em minus 0.4em\relax
  North-Holland, 1973.

\bibitem{bermond1977line}
J.-C. Bermond, M.-C. Heydemann, and D.~Sotteau, ``Line graphs of hypergraphs
  {I},'' \emph{Discrete Mathematics}, vol.~18, no.~3, pp. 235--241, 1977.

\bibitem{chung1997spectral}
F.~Chung, \emph{Spectral graph theory}.\hskip 1em plus 0.5em minus 0.4em\relax
  American Mathematical Soc., 1997.

\bibitem{Devine2006}
K.~D. Devine, E.~G. Boman, R.~T. Heaphy, R.~H. Bisseling, and U.~V. Catalyurek,
  ``Parallel hypergraph partitioning for scientific computing,'' ser.
  IPDPS.\hskip 1em plus 0.5em minus 0.4em\relax IEEE, 2006, p. 124.

\bibitem{Estrada2006}
E.~Estrada and J.~A. Rodr{\'{i}}guez-Vel{\'{a}}zquez, ``Subgraph centrality and
  clustering in complex hyper-networks,'' \emph{Physica A: Statistical
  Mechanics and its Applications}, vol. 364, pp. 581--594, 2006.

\bibitem{FeSHeE20}
S.~Feng and et. al, ``Hypergraph models of biological networks to identify
  genes critical to pathogenic viral response,'' \emph{BMC Bioinformatics},
  vol.~22, no.~1, p. 287, 2021.

\bibitem{Fiedler1973}
M.~Fiedler, ``\BIBforeignlanguage{eng}{Algebraic connectivity of graphs},''
  \emph{\BIBforeignlanguage{eng}{Czechoslovak Mathematical Journal}}, vol.~23,
  no.~2, pp. 298--305, 1973.

\bibitem{goh2007human}
K.-I. Goh, M.~E. Cusick, D.~Valle, B.~Childs, M.~Vidal, and A.-L. Barab{\'a}si,
  ``The human disease network,'' \emph{Proceedings of the National Academy of
  Sciences}, vol. 104, no.~21, pp. 8685--8690, 2007.

\bibitem{gustavson_1978_two}
F.~G. Gustavson, ``Two fast algorithms for sparse matrices: Multiplication and
  permuted transposition,'' \emph{ACM Trans. Math. Softw.}, vol.~4, no.~3, p.
  250–269, 1978.

\bibitem{heintz_mesh_2019}
B.~Heintz, R.~Hong, S.~Singh, G.~Khandelwal, C.~Tesdahl, and A.~Chandra,
  ``{MESH}: A flexible distributed hypergraph processing system,'' in
  \emph{2019 IEEE International Conference on Cloud Engineering (IC2E)}.\hskip
  1em plus 0.5em minus 0.4em\relax {IEEE}, 2019, pp. 12--22.

\bibitem{iacopini_simplicial_2019}
I.~Iacopini, G.~Petri, A.~Barrat, and V.~Latora, ``Simplicial models of social
  contagion,'' \emph{Nature Communications}, vol.~10, no.~1, p. 2485, 2019.

\bibitem{imdb_interfaces}
\BIBentryALTinterwordspacing
{IMDB Interfaces}. [Online]. Available: \url{https://www.imdb.com/interfaces/}
\BIBentrySTDinterwordspacing

\bibitem{tbbrepo}
\BIBentryALTinterwordspacing
{Intel Threading Building Blocks (TBB)}, 2021. [Online]. Available:
  \url{https://github.com/oneapi-src/oneTBB}
\BIBentrySTDinterwordspacing

\bibitem{jaja_intro_1992}
J.~Jaja, \emph{An Introduction to Parallel Algorithms}.\hskip 1em plus 0.5em
  minus 0.4em\relax Addison-Wesley, 1992.

\bibitem{javidian_hypergraph_2020}
M.~A. Javidian, Z.~Wang, L.~Lu, and M.~Valtorta, ``On a hypergraph
  probabilistic graphical model,'' \emph{Annals of Mathematics and Artificial
  Intelligence}, vol.~88, no.~9, pp. 1003--1033, 2020.

\bibitem{Jenkins_2018_chapel}
L.~{Jenkins}, T.~{Bhuiyan}, S.~{Harun}, C.~{Lightsey}, D.~{Mentgen}
  \emph{et~al.}, ``Chapel hypergraph library (chgl),'' in \emph{2018 IEEE High
  Performance extreme Computing Conference (HPEC)}, 2018, pp. 1--6.

\bibitem{jiang_hyperx_2019}
W.~Jiang, J.~Qi, J.~X. Yu, J.~Huang, and R.~Zhang, ``{HyperX}: A scalable
  hypergraph framework,'' \emph{{IEEE} Transactions on Knowledge and Data
  Engineering}, vol.~31, pp. 909 -- 922, 2019.

\bibitem{Karypis2000}
G.~Karypis and V.~Kumar, ``Multilevel k-way hypergraph partitioning,''
  \emph{{VLSI} Design}, vol.~11, no.~3, pp. 285--300, 2000.

\bibitem{KiS17}
S.~Kirkland, ``Two-mode networks exhibiting data loss,'' \emph{J Complex
  Networks}, vol. 6:2, pp. 297--316, 2017.

\bibitem{knuth1993stanford}
D.~E. Knuth, \emph{The Stanford GraphBase: a platform for combinatorial
  computing}.\hskip 1em plus 0.5em minus 0.4em\relax ACM, 1993, vol.~1.

\bibitem{kunegis2013konect}
J.~Kunegis, ``Konect: the koblenz network collection,'' in \emph{Proceedings of
  the 22nd Intl. Conference on World Wide Web}, 2013, pp. 1343--1350.

\bibitem{ActiveDNSdataset}
\BIBentryALTinterwordspacing
A.~Lab, ``{Active DNS project},'' 2020. [Online]. Available:
  \url{https://activednsproject.org/}
\BIBentrySTDinterwordspacing

\bibitem{LaNReJ20}
N.~W. Landry and J.~G. Restrepo, ``The effect of heterogeneity on hypergraph
  contagion models,'' \emph{Chaos: An Interdisciplinary Journal of Nonlinear
  Science}, vol.~30, no.~10, p. 103117, 2020.

\bibitem{leskovec2016snap}
J.~Leskovec and A.~Krevl, ``{SNAP} datasets: {S}tanford large network dataset
  collection; 2014,'' \emph{http://snap. stanford. edu/data}, 2016.

\bibitem{lewis2020centre}
R.~Lewis, ``Who is the centre of the movie universe? using python and networkx
  to analyse the social network of movie stars,'' \emph{CoRR}, vol.
  abs/2002.11103, 2020.

\bibitem{firoz_2020_efficient}
X.~T. Liu, J.~Firoz, and et. al, ``Parallel algorithms for efficient
  computation of high-order line graphs of hypergraphs,'' in \emph{Proc. of the
  28th IEEE Intl Conference on High Performance Computing, Data, and Analytics
  (HiPC)}.\hskip 1em plus 0.5em minus 0.4em\relax IEEE, 2021, p. In press.

\bibitem{marinov2016practical}
M.~Marinov, N.~Nash, and D.~Gregg, ``Practical algorithms for finding extremal
  sets,'' \emph{Journal of Experimental Algorithmics (JEA)}, vol.~21.

\bibitem{MiM00}
M.~Minas, ``Hypergraphs as a uniform diagram representation model,'' ser.
  TAGT'98.\hskip 1em plus 0.5em minus 0.4em\relax Springer-Verlag, 1998, p.
  281–295.

\bibitem{SpGEMMrepo}
Y.~Nagasaka, S.~Matsuoka, A.~Azad, and A.~Bulu{\c{c}}, ``{Sparse General
  Matrix-Matrix Multiplication for multi-core CPU and Intel KNL },'' 2020.

\bibitem{naik2018intersection}
R.~N. Naik, ``On intersection graphs of graphs and hypergraphs: A survey,''
  \emph{arXiv preprint arXiv:1809.08472}, 2018.

\bibitem{structure_newman_2001}
M.~E.~J. Newman, ``The structure of scientific collaboration networks,''
  \emph{Proceedings of the National Academy of Sciences of the United States of
  America}, vol.~98, no.~2, pp. 404--409, 2001.

\bibitem{amazon_data}
J.~Ni, J.~Li, and J.~McAuley, ``Justifying recommendations using
  distantly-labeled reviews and fine-grained aspects,'' in \emph{Proceedings of
  the 2019 Conference on Empirical Methods in Natural Language Processing and
  the 9th International Joint Conference on Natural Language Processing
  (EMNLP-IJCNLP)}, 2019, pp. 188--197.

\bibitem{PaAPeG17}
A.~Patania, G.~Petri, and F.~Vaccarino, ``The shape of collaborations,''
  \emph{EPJ Data Science}, vol.~6, no.~1, p.~18, 2017.

\bibitem{pinero2020disgenet}
\BIBentryALTinterwordspacing
J.~Pi{\~n}ero and et. al, ``The disgenet knowledge platform for disease
  genomics: 2019 update,'' \emph{Nucleic acids research}, vol.~48, no.~D1, pp.
  D845--D855, 2020. [Online]. Available: \url{https://www.disgenet.org}
\BIBentrySTDinterwordspacing

\bibitem{shun2020practical}
J.~Shun, ``Practical parallel hypergraph algorithms,'' in \emph{Proc. of the
  25th ACM SIGPLAN Symposium on Principles and Practice of Parallel
  Programming}, 2020, pp. 232--249.

\bibitem{austin_data}
N.~Veldt, A.~R. Benson, and J.~Kleinberg, ``Minimizing localized ratio cut
  objectives in hypergraphs,'' in \emph{Proceedings of the 26th {ACM} {SIGKDD}
  International Conference on Knowledge Discovery and Data Mining}.\hskip 1em
  plus 0.5em minus 0.4em\relax {ACM} Press, 2020.

\bibitem{zien_multilevel_1999}
J.~Y. Zien, M.~D. Schlag, and P.~K. Chan, ``Multilevel spectral hypergraph
  partitioning with arbitrary vertex sizes,'' \emph{IEEE Transactions on
  Computer-Aided Design of Integrated Circuits and Systems}, vol.~18, no.~9,
  pp. 1389--1399, 1999.

\end{thebibliography}

\end{document}